\documentclass[preprint,10pt,aps, numbers,sort&compress,amsmath,amssymb,amsfonts]{elsarticle}

\usepackage{amssymb}
\usepackage{tikz}
\usepackage{amsmath}
\usepackage{hyperref}
\usepackage{alltt}
\usepackage{slashed}

\usepackage{setspace}
\usepackage{color}

\usepackage{geometry}
\journal{CPC}

\begin{document}

\newcommand{\mhalfo}{\frac{1}{2}}	
\newcommand{\mhalf}[1]{\frac{#1}{2}}
\newcommand{\ka}{\kappa}
\newcommand{\al}{\alpha}
\newcommand{\be}{\beta}
\newcommand{\ga}{\gamma}
\newcommand{\la}{\lambda}
\newcommand{\de}{\delta} 
\newcommand{\vp}[0]{\varphi} 
\newcommand{\vpb}[0]{\bar{\varphi}} 
\newcommand{\equ}[1]{\begin{equation} #1 \end{equation}}
\newcommand{\ba}{\begin{align}}
\newcommand{\ea}{\end{align}}	
\newcommand{\eref}[1]{Eq.~(\ref{#1})}
\newcommand{\fref}[1]{Fig.~\ref{#1}}
\newcommand{\ddotp}[1]{\frac{d^d #1}{(2\pi)^d}}	
\newcommand{\nnnl}{\nonumber\\}	
\newcommand{\G}[1]{\Gamma(#1)}
\newcommand{\nq}{\nu_1}	
\newcommand{\nw}{\nu_2}	
\newcommand{\nd}{\nu_3}	
\newcommand{\dhalf}{\frac{d}{2}} 
\newcommand{\fig}[5]{\begin{figure}[#1]\centering\includegraphics[#5]{#3}\caption{#4}\label{#2}\end{figure}}
\newcommand{\cb}{$\bigstar$} 
\newcommand{\ce}{$\blacksquare$} 
\newcommand{\tw}[1]{\texttt{#1}} 
\newcommand{\beq}{\begin{eqnarray}}
\newcommand{\eeq}{\end{eqnarray}}
\newcommand{\str}{{\rm STr}}
\newcommand{\hier}{ {\color{red}\rule{1.0\linewidth}{1pt}}}
\newcommand{\colM}[1]{{\color{blue}{#1}}}
\newcommand{\colMT}[1]{{\color{magenta}{#1}}}
\newcommand{\colD}[1]{{\color{red}{#1}}}
\newcommand{\colF}[1]{{\color{green}{#1}}}
\newcommand{\Mathematica}{\textit{Mathematica}}
\newcommand{\DoFun}{\textit{DoFun}}
\newcommand{\DoDSERGE}{\textit{DoDSERGE}}
\newcommand{\DoAE}{\textit{DoAE}}
\newcommand{\DoFR}{\textit{DoFR}}

\begin{frontmatter}



\title{Algorithmic derivation of functional renormalization group equations and Dyson-Schwinger equations}


\author{Markus Q. Huber}
\ead{markus.huber@tu-darmstadt.de}
\address{Theoretisch-Physikalisches Institut, Friedrich-Schiller-Universit\"at Jena, Max-Wien-Platz~1, 07743 Jena, Germany\\
 Institut f\"ur Kernphysik, Technische Universit\"at Darmstadt, Schlossgrabenstr. 9, 64289 Darmstadt, Germany}
\author{Jens Braun}
\ead{j.braun@uni-jena.de}

\address{Theoretisch-Physikalisches Institut, Friedrich-Schiller-Universit\"at Jena, Max-Wien-Platz~1, 07743 Jena, Germany}

\begin{abstract}
We present the \textit{Mathematica} application \textit{DoFun}\footnote{The application is 
available from \url{http://theorie.ikp.physik.tu-darmstadt.de/~mqh/DoFun}.} 
which allows to derive Dyson-Schwinger equations and renormalization group flow equations for $n$-point functions in a simple manner. 
\DoFun\ offers several tools which considerably simplify the derivation of these equations from a given physical action.
We discuss the application of \DoFun\ by means of two different types of quantum field theories, namely
a bosonic $O(N)$ theory and the Gross-Neveu model.
\end{abstract}

\begin{keyword}
Dyson-Schwinger equations \sep functional renormalization group equations \sep correlation functions \sep quantum field theory


\end{keyword}

\end{frontmatter}

{\bf PROGRAM SUMMARY}

\begin{small}
\medskip 
\noindent
{\em Program Title:} DoFun                                          \\
{\em Version number:} 2.0.0\\
{\em Licensing provisions:} CPC non-profit use license                                  \\
{\em Programming language:}  Mathematica 7 and higher                               \\
{\em Operating system:} all on which Mathematica is available (Windows, Unix, MacOS)                                       \\
{\em PACS:} 11.10.-z,03.70.+k,11.15.Tk                                                 \\
\noindent
{\em Nature of problem:} Derivation of functional renormalization group equations and Dyson-Schwinger equations from the action of a given theory.\\
{\em Unusual features:} The results can be plotted as Feynman diagrams in Mathematica. The output is compatible with the syntax of many other programs and is therefore suitable for further (algebraic) computations.\\
\end{small}

\section{Introduction}
The derivation of Dyson-Schwinger equations (DSEs) 
around 1950~\cite{Schwinger:1951ex,Schwinger:1951hq,Dyson:1949ha} and renormalization 
group equations (RGEs) in the early 
1970s~\cite{Callan:1970yg,Symanzik:1970rt,Wilson:1971bg,Wilson:1971dh,Wilson:1973jj,Wegner:1972ih} has
equipped us with powerful tools for an analysis of the dynamics of
quantum field theories. Both approaches have been further developed in the past 30 years. In fact,
many formulations of these two methods now rely on a formulation in terms of
so-called generating functionals for Green functions~\cite{Itzykson:1980ft,Polchinski:1983gv,Wetterich:1992yh,Liao:1992fm,Morris:1993qb}.

These days functional approaches, such as DSEs, RGEs or the $n$-PI formalism, see e.~g.~\cite{Berges:2004pu,Pawlowski:2005xe,Carrington:2010qq,Carrington:2009kh}, 
are well-established for studies of quantum field theories.
Apart from functional approaches, Monte-Carlo simulations based on a discretized action have been 
extensively used to study non-perturbative phenomena. In fact, so-called QCD lattice simulations are 
currently the most powerful tool available for a study of full QCD. However, the implementation
of fermions in such simulations continues to be a non-trivial task. Functional approaches
are also non-perturbative but do not have problems arising from a discretized action or from the 
implementation of fermionic degrees of freedom. 
However, the application of DSEs and RGEs eventually requires in most cases a truncation of the
full system of equations of a given theory. From this point of view, it is clear that 
Monte-Carlo simulations and functional approaches are 
complementary approaches for studies of non-perturbative phenomena in quantum field theories.

DSEs and non-perturbative RGEs have been successfully employed to gain a better understanding
of a large and diverse variety of quantum field theories. For instance, detailed studies of condensed-matter 
systems,
see e.~g.~\cite{Honerkamp1,Honerkamp2,Honerkamp3}, 
critical phenomena, 
see e.~g.~\cite{Tetradis:1993ts,Litim:2002cf,Braun:2008sg,Benitez:2009xg,Litim:2010tt}, 
few- and many-body physics, 
see e.~g.~\cite{Diehl:2005ae,Diehl:2007th,Diehl:2007xz,Diehl:2009ma,Bartosch:2009zr,Krippa:2009vu}, 
gravity, 
see e.~g.~\cite{Reuter:1996cp,Litim:2003vp,Codello:2006in,Machado:2007ea,Eichhorn:2010tb}, 
QCD, 
see e.~g.~\cite{Aguilar:2008xm,Aguilar:2006gr,Alkofer:2008jy,Alkofer:2008tt,
Alkofer:2008bs,Alkofer:2003jj,Alkofer:2000wg,Berges:1997eu,Boucaud:2008ky,Braun:2005uj,Braun:2006jd,Braun:2008pi,
Braun:2009gm,
Campagnari:2010wc,Epple:2006hv,Fischer:2002hna,Fischer:2006ub,Fischer:2009tn,
Fischer:2008uz,Fischer:2009gk,Fischer:2004uk,Fischer:2003rp,Fischer:2009wc,
Fischer:2006vf,Gies:2002hq,Herbst:2010rf,Huber:2009wh,Huber:2009tx,
Jungnickel:1995fp,Kondo:2010ts,Maas:2004se,Maas:2005hs,Pawlowski:2003hq,vonSmekal:1997vx,Schleifenbaum:2006bq,
Skokov:2010uh,Watson:2010cn,Watson:2006yq}, 
standard model physics, 
see e.~g.~\cite{Gies:2003dp,Scherer:2009wu}, and 
supersymmetry,
see e.~g.~\cite{Falkenberg:1998bg,Synatschke:2008pv,Synatschke:2009nm,Synatschke:2010ub},
are available these days. Furthermore, DSEs are also used as an alternative to the traditional Feynman graph 
approach in perturbation theory, see for example \cite{Caravaglios:1995cd,Kanaki:2000ey,Mangano:2002ea}. 
For reviews on and introductions to the application of DSEs 
and non-perturbative RGEs we refer to
e.~g.~\cite{Alkofer:2000wg,Fischer:2006ub,Binosi:2009qm,Roberts:1994dr}
and~\cite{Litim:1998nf,Bagnuls:2000ae,Berges:2000ew,Polonyi:2001se,Delamotte:2003dw,Pawlowski:2005xe,Gies:2006wv,%
Delamotte:2007pf,Sonoda:2007av,Rosten:2010vm,Kopietz:2010zz}, 
respectively.

The applicability of DSEs and RGEs to very different theories is indeed an attractive feature of these approaches.
However, the intricacy of the derivation of DSEs and RGEs scales non-linearly with the complexity 
of the theory.  
Therefore pushing a computation to a higher 
level of accuracy often requires a big effort as the number of terms increases considerably. Consequently, 
tools facilitating the derivation of such equations are helpful for future studies with DSEs and 
non-perturbative RGEs and our {\it Mathematica} \cite{Wolfram:2004} program \textit{DoFun}\footnote{Short for "Derivation of Functional equations".} does exactly 
that: It allows for an automatic derivation of DSEs and non-perturbative RGEs from a given action. Of course, finding a suitable 
ansatz for the effective action for a given problem is left to the user and remains to be the most difficult step
from a physical point of view. 

The program \textit{DoFun}
is a further development of the {\it Mathematica} \textit{DoDSE} package\footnote{Short for "Derivation of Dyson-Schwinger Equations".} \cite{Alkofer:2008nt} which was restricted to a
derivation of DSEs in symbolic form. There are two new main features: First, the derivation of RGEs is 
now included and, second, the symbolic results can be transformed into the corresponding algebraic expressions. 
For the latter task two additional packages were added to \textit{DoDSE}: \textit{DoAE} 
and \textit{DoFR}\footnote{The names are abbreviations of "Derivation of Algebraic Expressions" 
and "Derivation of Feynman Rules", respectively.}. To account for the inclusion of RGEs the former package \textit{DoDSE} 
was renamed \textit{DoDSERGE}. These three packages form the content of the program \textit{DoFun}.

In the process of the development of \textit{DoFun} we always 
had in mind that there indeed exists a variety of programs to deal with single steps of the 
derivation of functional equations like performing Dirac traces or simplifications of 
the color structure. 
We did not intend to force the user to learn an additional new program syntax for these kinds
of tasks but to open up the possibility of using our program in combination with 
the corresponding available programs. 
Therefore, our goal was to stay as general as possible and to allow for the combination of \DoFun\ 
with many well-established programs, such as \textit{TRACER} \cite{Jamin:1991dp} or \textit{FeynCalc} \cite{Mertig:1990an}. 
The program \textit{DoFun} only performs the most basic simplifications and the output can then be handled
with other programs. For example, the user may want to stick to his favorite program to deal with
the color algebra or the user may want to use his own programs for certain operations.
The latter is often very efficient, since the code can then be 
specifically tailored for the problem at hand.
Following this general approach allows a high flexibility for the user and the treatment of a 
huge variety of theories. 

We are not aware of the existence of other programs for the derivation of DSEs 
whereas other packages for the derivation of functional RGEs indeed exist, 
see Refs.~\cite{Haas:2008,FisterHaasPawlowski,Benedetti:2010nr}.
However, as mentioned above, \textit{DoFun} is based on the application \textit{DoDSE} and exploits 
similarities in the derivation of DSEs and RGEs. It thus can be viewed as complementary to other existing programs.

The aim of \textit{DoFun} is \textit{not} to replace the manual derivation process completely, 
but to provide additional help in cases where the manual derivation becomes too cumbersome. 
While $O(N)$ models may still be more easily accessible with pen and paper, studies of more complicated theories 
may benefit in various ways from \DoFun:
\begin{itemize}
 \item In general, a derivation of functional equations by hand becomes tedious when several different fields 
or large numbers of operators are taken into account.
 \item With a package for an automatized derivation of functional equations, equations at the next higher level 
of the truncation are easily accessible. For example, this is particularly useful to explore the role of higher-order operators.
 \item Theories with complicated tensor structures benefit from the connection to a computer algebra system 
as tensors can be directly computed or simplified.
 \item Finally, the graphical output of \textit{DoFun} is often helpful for an illustration of 
the basic structure of the equations.
\end{itemize}

\DoFun\ is an extension of \textit{DoDSE}. Both have already proven to be useful for a number of studies.
In fact, \textit{DoDSE} was first used for the derivation of the DSEs of the 
maximally Abelian gauge \cite{Huber:2009wh,Huber:2010ne}. Actually, the complexity of these equations was the reason for the development of \textit{DoDSE} 
in the first place. While a manual derivation of the equations would still have been possible but very ineffective, the computation 
of certain quantities would have been out of reach without the aid of a computer algebra system~\cite{Huber:2011mg}.
The advantages of an automatized derivation of equations has also been appreciated in the analysis of the Gribov-Zwanziger action, 
where one has to handle large expressions~\cite{Huber:2010cq,Huber:2010ne,Huber:2009tx}. Studies of scalar fields 
coupled to Yang-Mills theory benefited from \textit{DoDSE} as well \cite{Fister:2010ah,Fister:2010yw,Alkofer:2010tq,Macher:2010dt}. 
Recently, \DoFun\ has been used for an investigation of bound states appearing in the BRST quartets of QCD \cite{Alkofer:2011pe,Alkofer:2011uh}.

\begin{figure}[tb]
 \begin{center}
  \includegraphics[width=\textwidth]{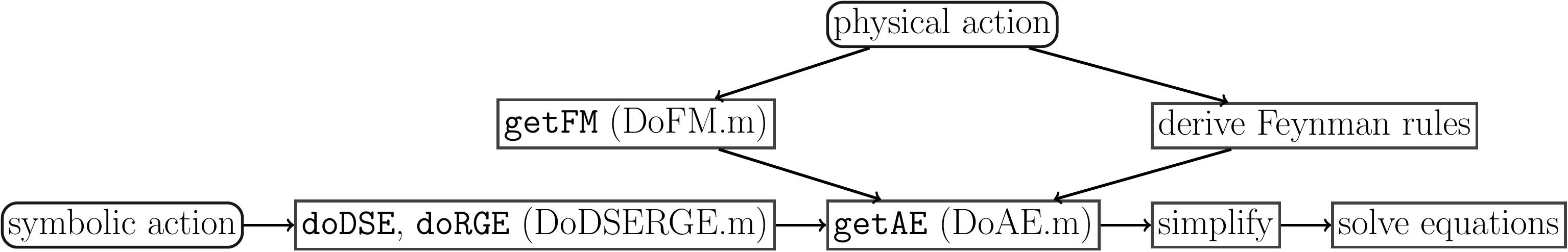}
  \caption{\label{fig:workflow}Workflow of \textit{DoFun}. 
}
 \end{center}
\end{figure}

Working with \textit{DoFun} involves several steps: The first one is the derivation of the equations in symbolic 
form using the functions \tw{doDSE} or \tw{doRGE}. As input they require the action in symbolic form. 
In order to transform the results 
into algebraic expressions with \textit{DoAE}, Feynman rules have to be defined. They can be either derived by hand from the physical 
action or  with the aid of other available packages, such as~\textit{DoFR}. The algebraic expressions represent the final 
output of \textit{DoFun}. Further manipulations are up to the user.
The generic workflow with \textit{DoFun} is summarized in Fig.~\ref{fig:workflow}.

To make \textit{DoFun} easily accessible for the user we implemented documentation directly into 
the \textit{Mathematica} help system. For quick reminders of the syntax of a function one 
can use the command~\tw{?}, e.~g.
\begin{verbatim}
?doRGE
\end{verbatim}
which also provides simple examples for many functions.
For more detailed information we also included a section into the \textit{Documentation Center} of \textit{Mathematica}. 
There, tutorials and guides introducing the basic features of \DoFun\ can be found. It can be accessed via \textit{Help} 
$\rightarrow$ \textit{Documentation Center} $\rightarrow$ \textit{Add-Ons and Packages} $\rightarrow$ \textit{DoFun} $\rightarrow$ \textit{Documentation}. 
All functions of \textit{DoFun} have a help entry which can be opened by moving the 
cursor into or behind the function name and pressing the key F1.

To install \DoFun\ the directory \texttt{DoFun} should be copied into the \textit{Mathematica} subdirectory \texttt{Applications}.\footnote{Under a Unix system this is typically \texttt{$\tilde{}$/.Mathematica/Applications}.}
Within \textit{Mathematica} the program \texttt{DoFun} can then be loaded with \texttt{<<DoFun$\grave{ }$}{ } or \texttt{Needs["DoFun$\grave{ }$"]} and the help system is accessible.

The paper is organized as follows: In Sect.~\ref{sec:FRGDSE} we briefly summarize the derivation of DSEs and
non-perturbative (functional) RGEs. This part can be skipped by readers familiar with the topic. In Sect.~\ref{sec:program} the basic constructs 
and quantities of the program \DoFun\ are introduced. In Sect.~\ref{sec:programDetails} we explain the main functions of \DoFun. 
Their usage is then demonstrated in Sect.~\ref{sec:examples} by deriving the flow equations of two substantially different
theories, namely an $O(N)$ model in $d$ space-time dimensions and the Gross-Neveu model in $d=3$ space-time
dimensions. Comments concerning the implementation of fermions are given in \ref{app:fermions}.
In \ref{app:versions} we give a short version summary of \DoFun.

\section{Functional RG equations and Dyson-Schwinger equations}\label{sec:FRGDSE}
\subsection{Actions and correlation functions}
The program \textit{DoFun} deals with the effective action $\Gamma[\Phi]$ as the underlying quantity for the
derivation of correlation functions. The field $\Phi$ denotes a collective field vector, e.~g. $\Phi=\{\varphi^1(x),\dots,\varphi^{N}(x)\}$
for an $O(N)$ model or $\Phi=\lbrace A^a_\mu(x), c^a(x), \bar{c}^a(x), q^i(x), \bar{q}^i(x) \rbrace$ for QCD.
Here and in the following we shall assume that the index of the field $\Phi$ refers to the field  type, the momentum and to 
all indices associated with internal symmetry groups (except if these indices are stated explicitly). 
The effective action is defined as follows:
\begin{align}
 \Gamma[\Phi]&:=\sup_J(-W[J]+\Phi_i J_i )\,,
\end{align}
where the $J_i$'s denote sources for the fields $\Phi_i$. Note that we sum/integrate over all possible values of 
index variables that appear twice in a single term.
On the other hand, the generating functional $W[J]$ is related to the bare action $S[\phi]$ via the path integral:
\begin{align}
Z[J]=\int D[\phi] e^{-S + \phi_j J_j}=:e^{W[J]}.\label{eq:pathint}
\end{align}
Here, the $\phi_i$'s denote the quantum fields. The so-called average field $\Phi$ is given by
\begin{equation}
\Phi_{i}\equiv \left\langle \phi_{i}\right\rangle _{J}=\frac{\delta W}{\delta J_{i}}=Z[J]^{-1}\int D[\phi] \phi_i e^{-S + \phi_j J_j} .
\end{equation}
The physical expectation value of the fields $\phi_i$ is then obtained by setting 
the external source $J$ to zero, 
$\Phi_{\rm phys}:=\left\langle \phi_{i}\right\rangle _{J=0}$.

The effective action can be expanded about, e.~g., the physical ground state $\Phi_{\rm phys}$ as follows:
\begin{equation}
\Gamma[\Phi]=\sum_{n=0}^{\infty}\,\frac{1}{\mathcal{N}^{i_1\ldots i_n}}\sum_{i_1\ldots i_n} \Gamma^{i_1\ldots i_n} 
(\Phi_{i_1}-\Phi_{i_1,{\rm phys}})\ldots(\Phi_{i_n}-\Phi_{i_n,{\rm phys}})\,,
\end{equation}
where $\mathcal{N}^{i_1 \ldots i_n}$ is the corresponding symmetry factor and 
the (physical) $n$-point functions are given by the expansion
coefficients\footnote{For convenience, we have inserted a minus sign appearing in our definition of the vertices~$\Gamma^{i_1\ldots i_n}$.
With this definition, the sign of 
any diagram can be obtained straightforwardly as it does not depend on the number of vertices in a diagram.}:
\begin{subequations}\label{eq:effActions}
\begin{align}
 \Gamma^{ij}:=&\Gamma^{ij}_{J=0}=\frac{\delta ^2 \Gamma[\Phi]}{\de \Phi_{i}\de \Phi_{j}}\Bigg|_{\Phi=\Phi_{\rm phys}},\\
 \Gamma^{i_1\ldots i_n}:=&\Gamma^{i_1\ldots i_n}_{J=0}=-\frac{\delta ^{n}\Gamma[\Phi]}{\de \Phi_{i_1}\ldots \de \Phi_{i_n}}\Bigg|_{\Phi=\Phi_{\rm phys}} \label{eq:vertexConvention}.
\end{align}
\end{subequations}
Note that the physical ground-state may even be space-time dependent as recently discussed 
in the context of Gross-Neveu and Nambu-Jona-Lasinio models, see e.~g. Refs.~\cite{Thies:2003kk,Basar:2008im,Nickel:2009wj,Kojo:2009ha}.  

From the generating functional $W[J]$ of connected correlation functions we obtain the propagator, 
i.~e. the inverse of the two-point function:
\begin{align}\label{eq:prop}
 D_J^{ij}:=\frac{\de W[J]}{\de J_i \de J_j}
=\left(\frac{\delta^{2}\Gamma[\Phi]}{\delta\Phi_{i}\delta\Phi_{j}}\right)^{-1}=\left( \Gamma_J ^{ij} \right)^{-1}\,,
\end{align}
where the index $J$ denotes the dependence of the two-point function on the source $J$.\footnote{Note that $\de \Gamma/\de \Phi_i=J_i$.}
From this it follows that the derivative of the propagator with respect to a field yields two propagators and 
a three-point function\footnote{Note that $\de (M M^{-1})/\de\Phi=0$.}. Overall, we only 
require the following set of differentiation rules:
\begin{subequations}\label{eq:derivatives}
\begin{align}
\frac{\delta}{\delta\Phi_{i}}D^{jk}_{J} & =\frac{\delta}{\delta\Phi_{i}}\left(\frac{\delta^{2}\Gamma}{\delta\Phi_{j}\delta\Phi_{k}}\right)^{-1}\nnnl
 &=-\left(\frac{\delta^{2}\Gamma}{\delta\Phi_{j}\delta\Phi_{m}}\right)^{-1}\left(\frac{\delta^{3}\Gamma}{\delta\Phi_{m}\delta\Phi_{i}\delta\Phi_{n}}\right)\left(\frac{\delta^{2}\Gamma}{\delta\Phi_{n}\delta\Phi_{k}}\right)^{-1}=D^{jm}_{J}\Gamma^{min}_{J}D^{nk}_{J}, \\
\frac{\delta}{\delta\Phi_{i}}\Phi_{j} &=\delta_{ij},\\
\frac{\delta}{\delta\Phi_{i}}\Gamma^{j_{1}\ldots j_{n}}_{J} & =-\frac{\delta\Gamma}{\delta\Phi_{i}\delta\Phi_{j_{1}}\ldots\delta\Phi_{j_{n}}}=\Gamma^{ij_{1}\ldots j_{n}}_{J}.\end{align}
\end{subequations}
These relations form our basis for an algorithm for the derivation of DSEs and functional RGEs which underlies \DoFun.

\subsection{Derivation of functional renormalization group equations}
\label{ssec:derivFRGEs}

Let us now briefly discuss the derivation of functional RG flow equations. Here, we shall follow the standard derivation
of the flow equation for the so-called effective average action given in Ref.~\cite{Wetterich:1992yh}.

First, we define the so-called effective average action $\Gamma_k$ which depends on 
a momentum-shell parameter $k$ and interpolates between the bare (classical) action and
the full quantum action $\Gamma$ for $k\to 0$. To this end, we introduce 
a so-called cutoff function $R_k$ into the path integral~\eqref{eq:pathint} as follows:
\beq
 S[\phi]\rightarrow S[\phi] + \Delta S_k [\phi] = S[\phi]+\frac1{2}\phi_i R^{ij}_k \phi_j\,.
\eeq
The so introduced (momentum) scale $k$ allows us to integrate out quantum fluctuations in a controlled way.
In momentum space the cutoff action $\Delta S_k [\phi]$ can be written as follows
\beq
 \frac1{2}\phi_i R_k^{ij} \phi_j&=&\frac1{2} \int \ddotp{q}\ddotp{q'} R_k^{ab}(q,q') \phi_a(q)\phi_b(q') \nnnl
 &=&\frac1{2} \int \ddotp{q} R_k^{ab}(q) \phi_a(q)\phi_b(-q)\,.
\eeq
Note that the regulator function $R_k$ is matrix-valued in field-space and has to obey the following
constraints:
\begin{itemize}
 \item[(1)] $\lim_{q^2/k^2 \to 0} R_k >0\,$,
 \item[(2)] $\lim_{k^2/q^2 \to 0} R_k = 0\,$, 
 \item[(3)] $\lim_{k\to\Lambda \to \infty} R_k =\infty\,$.
\end{itemize}
The first constraint implements an IR regularization for the path integral. The second constraint 
ensures that the regulator vanishes for $k\to 0$. Thus, we recover the standard generating functional $Z$
defined in Eq.~\eqref{eq:pathint} for $k\to 0$. 

The inclusion of the cutoff action $\Delta S_k$ renders all generating functionals $k$-dependent. In particular,
the effective action now depends on the momentum scale~$k\sim q$. The scale-dependent effective action $\Gamma_k$,
the so-called effective average action, is defined via a modified Legendre transformation:
\beq
 \Gamma_k[\Phi]=-W_k[J]+J_i \Phi_i - \frac1{2}\Phi_i R^{ij}_k \Phi_j \label{eq:defavgamma}
\eeq
with
\beq\label{eq:def-Phi-k}
 \Phi_i=\frac{\delta W_k[J]}{\delta J_i}=\langle \phi_i \rangle_J.
\eeq
By construction, the effective average action $\Gamma_k$ includes all quantum fluctuations associated 
with momenta $p\gtrsim k$. The modified Legendre transformation together with the third
constraint for the regulator function ensures that $\Gamma _k$ reduces to the classical 
action $S$ for $k\to \Lambda$. On the other hand, we obtain the full quantum effective
action $\Gamma$ in the limit $k\to 0$.

From the effective average action $\Gamma _k$ we may derive scale-dependent correlation functions $\Gamma _k ^{i_1 \dots i_n}$ 
by differentiating $\Gamma _k$ with respect to the fields.\footnote{From here on we suppress the index $J$ as it should be clear from the context if an expression depends on $J\neq 0$.} Thus, the differentiation rules given in Eq.~(\ref{eq:derivatives}) also 
apply to the effective average action $\Gamma_k$.

Taking the derivative of Eq.~\eqref{eq:defavgamma} with respect to the scale $k$ we obtain the RG flow
equation for the effective action, the so-called Wetterich equation~\cite{Wetterich:1992yh}:
\begin{align}\label{eq:flowEq}
 \partial_k \Gamma_k[\Phi]=& \frac1{2} \left[\left(\Gamma_{k}[\Phi] + R_k \right)^{-1}\right]^{ji} \partial_k R_k^{ij}\nnnl
\equiv & \frac1{2}\str \left(\Gamma_{k}[\Phi] + R_k \right)^{-1} \partial_k R_k\,.
\end{align}
Clearly, the function $R_k$ implements an IR regularization for the momentum integrals and specifies the
details of the Wilsonian momentum-shell integrations. 
Note that the super trace $\str$ includes a minus sign for Grassmann-valued fields. 

The package \textit{DoDSERGE} for the derivation of the flow equations of $n$-point functions
 is based on a simple reformulation of Eq.~\eqref{eq:flowEq}:
\begin{align}\label{eq:flowEqLog}
 \partial_t \Gamma_k[\Phi]=& \frac1{2} \str\, \tilde{\partial}_t \ln \left(\Gamma_{k}[\Phi] + R_k \right),
\end{align}
where $t=\ln (k/\Lambda)$ with $\Lambda$ being a UV cutoff scale. The derivative $\tilde{\partial}_t$ only acts
on the regulator $R_k$. This formulation of the flow equation represents the starting point for the derivation 
of RG flow equations with \textit{DoFun}. The derivative $\tilde{\partial}_t$ is then taken at the end of the generation
of the flow equations. This last step increases the number of diagrams since it generates diagrams with the same topology
but with the regulator insertions attached to different internal lines.
Note that it is possible to suppress the derivative~$\tilde{\partial}_t$,
see Sect.~\ref{ssec:doRGE}. This is in fact useful if we are only interested in checking the structure of an equation.

The flow equations for, e. g., mass terms or so-called wave-function renormalizations 
are extracted from the flow equations of the $n$-point correlation functions. To obtain them, 
we take functional derivatives of Eq.~(\ref{eq:flowEqLog}) with respect to the fields. 
The first derivative yields
\begin{align}\label{eq:firstDer}
 \frac{\de}{\de \Phi_a}\partial_t \Gamma_k[\Phi]=& \frac1{2} \str
 \left\lbrace \tilde{\partial}_t \frac{\de ^3\Gamma_{k}}{\de \Phi_a \de \Phi_i \de \Phi_l }D^{lj}\right\rbrace.
\end{align}
where the indices $i$ and $j$ are contracted by the trace operation. 
From this, higher derivatives are then obtained by using the differentiation rules~\eqref{eq:derivatives}.
By definition, the derivative $\tilde{\partial}_t$ only acts on the propagators and inserts the derivative of a regulator into the loops:
\begin{align}\label{eq:dtD}
 \tilde{\partial}_t D^{ij} =-D^{il}(\partial_t R_k ^{lm})D^{mj}.
\end{align}

By taking derivatives with respect to the fields, the flow equations for the $n$-point functions 
can be derived in an exact form yielding an infinite  tower of coupled flow equations. Since it is impossible to study the flow of an effective action containing 
all operators allowed by the symmetries of a given theory, we have to restrict ourselves to that subspace of 
operators which we expect to be (most) relevant for the physical problem under consideration. 
This is the most difficult step since it requires a lot of physical insight into the problem 
in order to choose the correct subspace of operators. 

We now turn to a discussion of the command \texttt{doRGE} which performs the derivation of functional RGEs. It follows these steps:
\begin{enumerate}
 \item[(1)] Take the first derivative of $\partial _t \Gamma _k$, see Eq.~\eqref{eq:firstDer}.
 \item[(2)] Take higher derivatives using the command \texttt{derivRGE}. Note that this command
  takes into account the proper ordering of fermion fields.
 \item[(3)] The sources are set to their physical values with \texttt{setSourcesZeroRGE}.
 \item[(4)] Take the derivative with respect to $t=\ln(k/\Lambda)$, see Eq.~\eqref{eq:dtD}.
 \item[(5)] Identical Feynman diagrams are summed up with \texttt{identifyGraphsRGE}.
\end{enumerate}

\begin{figure}[tb]
 \begin{center}
  \includegraphics[width=\textwidth]{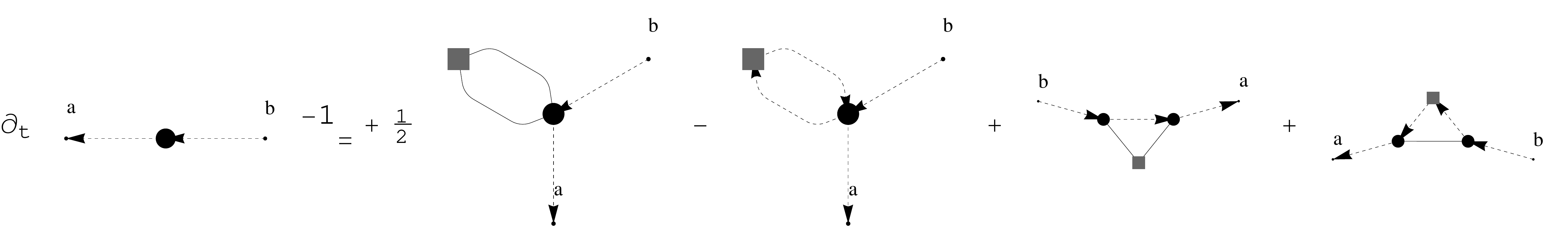}
  \caption{\label{fig:ghost2p} Ghost two-point function of Landau-gauge Yang-Mills theory. Solid
  lines denote gluons, dotted ones denote ghosts. The gray box represents the regulator insertion and blobs are 
  dressed vertices.}
 \end{center}
\end{figure}

Let us now give an explicit example of the steps performed by the command~\texttt{doRGE}. To this end,
we derive the flow equation for the ghost two-point function in Yang-Mills theory in the Landau gauge. 
There we have two types of fields, namely gluons and ghosts denoted by $A$ and $c$. The collective
field vector $\Phi$ is given by $\Phi=\{ A_\mu^a, c^a, \bar{c}^a \}$.
Note that the ghost fields $c$ and $\bar{c}$ are Grassmann-valued, i.~e. anti-commuting, fields. 
The extra rules required for their treatment are described in \ref{app:fermions}.
To obtain the ghost two-point function we first take the derivative of Eq.~\ref{eq:flowEqLog} %
with respect to a ghost and an anti-ghost field. This yields (step 1) 
\begin{align}
 \frac{\overset{\rightarrow}{\de}}{\de \bar{c}_a}&\partial_t \Gamma_k\frac{\overset{\leftarrow}{\de}}{\de c_b}=
  \frac{1}{2}\frac{\overset{\rightarrow}{\de}}{\de \bar{c}_a}\str \left\lbrace \tilde{\partial}_t \ln  \left(\Gamma _k^ {ij}+R_k ^{ij}\right)\right\rbrace\frac{\overset{\leftarrow}{\de}}{\de c_b}\nnnl
  &=\frac{1}{2}\str \, \tilde{\partial}_t\frac{\de^3 \Gamma _k}{\de \bar{c}_a \de \Phi_i \de \Phi_n} D_{nj} \frac{\overset{\leftarrow}{\de}}{\de c_b}.
\end{align}
The indices $a$ and $b$ are collective indices including the momenta and the color indices of the fields. On the other hand, the indices
$i$, $n$ and $j$ of the collective field $\Phi$ also include a field index which refers to the type of the field (gauge field, ghost, anti-ghost).
The open indices $i$ and $j$ are contracted by the trace operation but we perform the trace only at the end when we know if the corresponding fields are Grassmannian or not.
Performing the second derivative we obtain (step 2):
\begin{align}
 \frac{\overset{\rightarrow}{\de}}{\de \bar{c}_a}&\partial_t \Gamma_k\frac{\overset{\leftarrow}{\de}}{\de c_b}
 =\frac{1}{2}\str\, \tilde{\partial}_t \left( \left(\frac{\de ^3\Gamma _k}{\de \bar{c}_a \de \Phi_i \de \Phi_n} \frac{\overset{\leftarrow}{\de}}{\de c_b}\right) D^{nj}
 + \frac{\de^3 \Gamma _k}{\de \bar{c}_a \de \Phi_i \de \Phi_n} \left(D^{nj}\frac{\overset{\leftarrow}{\de}}{\de c_b}\right)\right)\nnnl
 =& \frac{1}{2}\str\, \tilde{\partial}_t \left\lbrace \frac{\de ^4\Gamma _k}{\de \bar{c}_a \de \Phi_i \de \Phi_n \de c_b}D^{nj}
 -\frac{\de ^3\Gamma _k}{\de \bar{c}_a \de \Phi_i \de \Phi_n}D^{nm}\frac{\de ^3 \Gamma _k}{\de \Phi_m \de \Phi_l \de c_b}D^{lj}\right\rbrace.
\end{align}
Before taking the (super-) trace we set the external sources to zero (step 3):
\begin{align}
 \partial_t \Gamma_{k,\bar{c}c}^{ab} 
 &=\frac{1}{2}\str\, \tilde{\partial}_t \Big\lbrace \frac{\de ^4\Gamma _k}{\de \bar{c}_a \de A_i \de A_n \de c_b}D_{AA}^{nj}
 +\frac{\de ^4\Gamma _k}{\de \bar{c}_a \de \bar{c}_i \de c_n \de c_b}D_{\bar{c}c}^{nj}+\frac{\de ^4\Gamma _k}{\de \bar{c}_a \de c_i \de \bar{c}_n \de c_b}D_{c\bar{c}}^{nj}\nnnl
 &-\frac{\de ^3\Gamma _k}{\de \bar{c}_a \de A_i \de c_n}D_{\bar{c}c}^{nm}\frac{\de ^3 \Gamma _k}{\de \bar{c}_m \de A_l \de c_b}D_{AA}^{lj}-\frac{\de ^3\Gamma _k}{\de \bar{c}_a \de c_i \de A_n}D_{AA}^{nm}\frac{\de ^3 \Gamma _k}{\de A_m \de \bar{c}_l \de c_b}D_{c\bar{c}}^{lj}
\Big\rbrace_{\Phi=0}\nnnl
 &=\frac{1}{2}\str\, \tilde{\partial}_t \Big\lbrace -\Gamma_{k,\bar{c}AAc}^{ainb}D_{AA}^{nj}-\Gamma_{k,\bar{c}\bar{c}cc}^{ainb}D_{\bar{c}c}^{nj}-\Gamma_{k,\bar{c}c\bar{c}c}^{ainb}D_{c\bar{c}}^{nj}\nnnl
 &\qquad\qquad\qquad -\Gamma^{k,\bar{c}Ac}_{ain}D_{\bar{c}c}^{nm}\Gamma_{k,\bar{c}Ac}^{mlb}D_{AA}^{lj} -\Gamma^{k,\bar{c}cA}_{ain}D_{AA}^{nm}\Gamma_{k,A\bar{c}c}^{mlb}D_{c\bar{c}}^{lj}
\Big\rbrace\nnnl
 &=\frac{1}{2}\str\, \tilde{\partial}_t \Big\lbrace -\Gamma_{k,\bar{c}AAc}^{ainb}D_{AA}^{nj}-\Gamma_{k,\bar{c}\bar{c}cc}^{ainb}D_{\bar{c}c}^{nj}-\Gamma_{k,\bar{c}\bar{c}cc}^{anib}D_{\bar{c}c}^{jn}\nnnl
 &\qquad\qquad\qquad -\Gamma^{k,\bar{c}Ac}_{ain}D_{\bar{c}c}^{nm}\Gamma_{k,\bar{c}Ac}^{mlb}D_{AA}^{lj} +\Gamma^{k,\bar{c}cA}_{ain}D_{AA}^{nm}\Gamma_{k,A\bar{c}c}^{mlb}D_{\bar{c}c}^{jl}
\Big\rbrace.\label{eq:ghglex}
\end{align}
In the last step we have brought the ghosts into canonical order. By setting $i=j$ we perform 
the trace. However, we have to take into account a minus sign if $i$ and $j$ correspond to indices of 
ghost fields:
\begin{align}
 \partial_t \Gamma_{k,\bar{c}c}^{ab}&=\frac{1}{2}\tilde{\partial}_t \Big( -\Gamma_{k,\bar{c}AAc}^{ainb}D_{AA}^{ni}+2\Gamma_{k,\bar{c}\bar{c}cc}^{ainb}D_{\bar{c}c}^{ni}
  -2\Gamma_{k,A\bar{c}c}^{ian}D_{\bar{c}c}^{nm}\Gamma_{k,A\bar{c}c}^{lmb}D_{AA}^{li}\Big).
\end{align}
By summing up the terms in Eq.~\eqref{eq:ghglex} we anticipated step~5. Finally we have to 
take the derivative of the propagators with respect to the RG time~$t$~(step~4):
\begin{align}\label{eq:ghost2pt}
 \partial_t \Gamma^{\bar{c}c}_{k,ab}&=\frac{1}{2}\Gamma_{k,\bar{c}AAc}^{ainb}D_{AA}^{nl}(\partial_t R_{k,AA}^{lm})D_{AA}^{mi}
 -\Gamma_{k,\bar{c}\bar{c}cc}^{ainb}D_{\bar{c}c}^{nl}(\partial_t R_{k,\bar{c}c}^{lm})D_{\bar{c}c}^{mi}\nnnl
 & +\Gamma_{k,A\bar{c}c}^{ian}D_{\bar{c}c}^{nm}\Gamma_{k,A\bar{c}c}^{lmb}D_{AA}^{lj}(\partial_t R_{k,AA}^{jo})D_{AA}^{oi}+\Gamma_{k,A\bar{c}c}^{ian}D_{\bar{c}c}^{nj}(\partial_t R_{k,\bar{c}c}^{jo})D_{\bar{c}c}^{om}\Gamma_{k,A\bar{c}c}^{lmb}D_{AA}^{li}.
\end{align}
This is the flow equation for the ghost two-point function. It is depicted in \fref{fig:ghost2p}. 
We observe that the pure ghost loop differs by a minus from the other diagrams as it should be.

\subsection{Derivation of Dyson-Schwinger equations}

For the sake of completeness we sketch the derivation of Dyson-Schwinger equations; 
details can be found in Ref.~\cite{Alkofer:2008nt}. For a short description of a graphical derivation we refer the reader to Ref. \cite{Alkofer:2008jy}. To this end, we start with the following integral of a total derivative 
which yields zero:
\begin{align}\label{eq:DSE-Z}
0=&\int D[\phi] \frac{\delta}{\delta \phi_i} e^{-S + \phi_j J_j}=\int D[\phi] \left( -\frac{\delta S}{\delta \phi_i} + J_i \right) e^{-S + \phi_j J_j}\nnnl
 =&\left( -\frac{\delta S}{\delta \phi'_i}\Bigg\vert_{\phi'_i=\delta/\delta J_i} +J_i \right) Z[J].
\end{align}
We are allowed to interchange integration and differentiation when we replace
the fields by derivatives with respect to the corresponding sources.
This yields the DSEs for full correlation functions. 
Next, we replace $Z[J]$ by $e^{W[J]}$ and multiply the resulting equation with $e^{-W[J]}$ from the left. Using
\begin{align}
e^{-W[J]}\left(\frac{\delta}{\delta J_i}\right)e^{W[J]}= \frac{\delta W[J]}{\delta J_i}+\frac{\delta}{\delta J_i}\,,
\end{align}
we find
\begin{align}
-\frac{\delta S}{\delta \phi_i}\Bigg\vert_{\phi_i=\frac{\delta W[J]}{\delta J_i}+\frac{\delta}{\delta J_i}} +J_i=0\,.
\end{align}
This is the (functional) DSE for connected correlation functions. 
To obtain the DSE for 1PI functions we perform a Legendre transformation of $W$ 
with respect to all sources. 
Using $\delta W[J]/\delta J_i=\Phi_i$ and 
\begin{align}
\frac{\delta}{\delta J_i}=\frac{\delta \Phi_j}{\delta J_i} \frac{\delta}{\delta \Phi_j}=\frac{\delta}{\delta J_i} \frac{\delta W}{\delta J_j} \frac{\delta}{\delta \Phi_j}=\frac{\delta^2 W}{\delta J_i \delta J_j} \frac{\delta}{\delta \Phi_j}=
D^{ij}_J \frac{\delta}{\delta \Phi_j}\,,
\end{align}
we are led to
\begin{align}\label{eq:DSE-master}
\frac{\delta \Gamma}{\delta \Phi_i}=\frac{\delta S}{\delta \phi_i}\Bigg\vert_{\phi_i=\Phi_i+D^{ij}_J  \, \delta/\delta \Phi_j}.
\end{align}
From this equation it is straightforward to derive DSEs for all 1PI functions by taking derivatives 
with respect to the fields. 
Recall that the summation over $j$ includes a sum over all fields of the theory. 
The command \texttt{doDSE} evaluates \eref{eq:DSE-master} and computes the $n$-point functions.
For explicit examples we refer the reader to Ref.~\cite{Alkofer:2008nt}.
\section{Basic usage of \textit{DoFun}}
\label{sec:program}
\subsection{Actions}
\label{ssec:actions}
The package \textit{DoDSERGE} represents the heart of~\textit{DoFun}. It contains the two 
commands~\tw{doDSE} and~\tw{doRGE} which allow the derivation of DSEs and RGEs, respectively. 
As input these commands use a symbolic form of the physical (effective) action. 
Contrary to the physical action written down with all indices and momenta, we shall refer to this symbolic form 
as symbolic action in the following. The symbolic action is a list of the types of fields and their interactions. 
This already suffices to determine the structure of the equations in terms of generic $n$-point functions and propagators. 

No specific information about the fields like their nature (e.~g. scalar, spinor or vector field) 
appear in the definition of the
symbolic action. The user only needs to provide the information whether a field is real, 
complex or Grassmann-valued. Details, such as the color or spin structure, only need to be provided when 
the package \textit{DoAE} is used. In fact, the command~\tw{getAE} allows to
generate specific expressions from the generic correlation functions obtained from~\tw{doDSE} and~\tw{doRGE}, respectively.
The resulting expressions can then be simplified by, e.~g., contracting indices. 
\DoFun\ offers only a tool to perform the simplest kind of contractions, namely those involving Kronecker deltas.
The implementation of an advanced handling of indices such as Lorentz or color indices is left to the user. 
However, we would like to stress that programs for these tasks exist and can be applied to the output of~\DoFun.

While the starting point for both \texttt{doDSE} and \texttt{doRGE} is an action, there is an important difference: 
\texttt{doDSE} requires the microscopic action $S$, whereas \texttt{doRGE} is based on an ansatz for the effective average action $\Gamma _k$. 
The latter is related to the classical action by $k\to\Lambda$: $\Gamma_{k\to\Lambda} \to S$.

As a simple example we consider a $\vp^4$ theory. The symbolic action for this theory reads
\begin{alltt}
 actionS=\(\{\{\vp,\vp\}, \{\vp,\vp,\vp,\vp\}\}\)
\end{alltt}
This action represents the bare action
\beq
 S[\vp] &=& \frac{1}{2}\int\ddotp{q}\vp(q)q^2 \vp(-q)\nnnl
 &&\qquad +\frac{1}{4!}\int \ddotp{q}\ddotp{r}\ddotp{s}\lambda\,
\vp(q)\vp(r)\vp(s)\vp(-q-r-s)
\eeq
for \texttt{doDSE} with $\lambda$ being the coupling constant. For \texttt{doRGE}, the symbolic 
action~\texttt{actionS} represents the ansatz for the effective average action
\begin{align}
 &\Gamma_k[\varphi]=\int \frac{1}{2}\ddotp{q}\vp(q) Z_k(q^2) q^2 \vp(-q)\\
 &\quad +\frac{1}{4!}\int \ddotp{q}\ddotp{r}\ddotp{s}\lambda_k(q,r,s,-q-r-s)\vp(q)\vp(r)\vp(s)\vp(-q-r-s)\nonumber
\end{align}
where $Z_k(q^2)$ is the so-called 
wave-function renormalization and $\lambda_k$ is a momentum-dependent coupling constant. 
Note that in the definition of the symbolic action \texttt{actionS} no signs or numerical factors appear. 
The program computes the latter from the multiplicity of the fields and their statistics. 
The bosonic or fermionic nature in this simple example is 
determined by the two-point function: if the list entries of length two have twice the same entry, 
e.g. $\{\vp,\vp\}$ in the example above, the field is considered to be bosonic. 
If there are two different fields, e.~g. $\{c,\bar{c}\}$, then they represent 
as Grassmann-valued fields where $\bar{c}$ is the anti-field of $c$. 

There are two cases when we have to override this rule: (a) the fields mix at the two-point level 
as it is the case, for instance, in the Gribov-Zwanziger formalism \cite{Zwanziger:1989mf}, (b) 
we have complex bosonic fields. 
This can be done with the option~\texttt{specificFieldDefinitions}, see below for a specific example.

We would like to stress that the action \tw{actionS} can also stand for a different theory, for example,
\beq
 S[\vp]&=& \frac{1}{2}\sum_i\int\ddotp{q}\vp^i(q)q^2 \vp^i(-q)\nnnl
  && \;\; +\frac{1}{8}\sum_{i,j}\int \ddotp{q}\ddotp{r}\ddotp{s}\lambda\,
\vp^i(q)\vp^i(r)\vp^j(s)\vp^j(-\!q\!-\!r\!-s).
\eeq
This ambiguity of the symbolic action is due to the fact that the combinatorics
of the equations can be the same for various theories.

As we use an ansatz for the effective average action in the derivation of RGEs, we may include operators
which are not present in the classical action $S$. These operators may be generated in the RG flow due to
quantum fluctuations. In any case, the initial conditions for the RG flow equations are chosen such
that $\Gamma_k \to S$ for $k\to\Lambda$. Thus, the ansatz for the effective average action
together with the initial conditions determine the quantum field theory under consideration. 
For instance, we may use the following symbolic action for a one-component scalar field theory
which is invariant under the transformation $\vp\rightarrow-\vp$:
\begin{alltt}
 \(\{\{\vp,\vp\},\{\vp,\vp,\vp,\vp\}\}\)
\end{alltt}
However, we may also use
\begin{alltt}
 \(\{\{\vp,\vp\},\{\vp,\vp,\vp,\vp\},\{\vp,\vp,\vp,\vp,\vp,\vp\}\}\)
\end{alltt}
The initial value of the six-boson interaction is then set to zero at the UV scale $k=\Lambda$ 
to ensure $\Gamma_k\to S$ for $k\to\Lambda$. Alternatively, the action under consideration can 
be specified by the maximal order in a given type of field,~e.~g.
\begin{alltt}
\(\{\{\vp,4\}\}\)
\end{alltt}
and
\begin{alltt}
\(\{\{\vp,6\}\}\)
\end{alltt}
respectively. To check this we use the command \texttt{generateAction} which generates the corresponding
action. For
\begin{alltt}
generateAction[\(\{\{\vp, 4\}\}\)]
\end{alltt}
we find that the action consists of three terms, namely a two-, three- and four-point 
function. We can enforce the reflection symmetry $\vp\rightarrow -\vp$ by adding the argument \textit{even}:
\begin{alltt}
 generateAction[\(\{\{\vp, 4,\) even\(\}\}\)]
\end{alltt}
Only vertices with an even number of legs are now included in the action. 

Let us now discuss complex scalar fields. To define the symbolic action, we have to use the 
option~\texttt{specificFieldDefinitions}. Using
\begin{alltt}
 actionCS=\(\{\{\vp,\bar{\vp}\}, \{\bar{\vp},\bar{\vp},\vp,\vp\}\}\)
\end{alltt}
the commands \texttt{doDSE} and \texttt{doRGE} assume that $\vp$ is a Grassmann-valued field by default.
To avoid this, we use the option \texttt{specificFieldDefinitions} in the derivation of 
functional equations. For the present case it reads
\begin{alltt}
  specificFieldDefinitions -> \(\{\vp, \bar{\vp}\}\)
\end{alltt}
and is given as an argument to \texttt{doDSE} or \texttt{doRGE}. An example is provided in Section~\ref{ssec:derDSEs}.

For theories in which fields mix at the level of the two-point function, 
such as the local Gribov-Zwanziger action \cite{Zwanziger:1989mf}, 
we refer to a tutorial provided in the \textit{Documentation Center} of \textit{Mathematica}. Indeed \DoFun\ has already been applied 
successfully to such types of theories~\cite{Huber:2010ne,Huber:2009tx,Huber:2010cq}.

In order to avoid confusion which types of ans\"atze for the effective average action, viz. which expansions, can be used with \DoFun\ we want to state explicitly that all ans\"atze are suitable as long as $n$-point functions can be defined from them. This includes also the usual derivative expansion.

\subsection{The basic constructs of \textit{DoFun}}
\label{ssec:language}

\textit{DoFun} uses the command \texttt{generateAction} to translate symbolic actions 
given in list form into its 'internal' language. The basic quantity of this language 
is the \texttt{op} construct. It acts as a container for fields, propagators, 
regulator insertions and vertices and as such it is 
responsible for keeping order in the expressions. \texttt{op} constructs also represent
the summation/integration over indices and they are indispensable for the
implementation of anti-commuting fields. The most general expression appearing in \textit{DoFun} consists of a sum of \tw{op} 
functions with an arbitrary complex-valued number as coefficient\footnote{Note that fields, vertices, regulator insertions or propagators
are not permitted as coefficients but only inside \tw{op} functions.}. 
Any numerical factors inside an \tw{op} operator are immediately put in front of it. Before we discuss more specific 
examples, we introduce fields, propagators, regulator insertions and vertices.

\begin{figure}[tb]
 \begin{center}
  \includegraphics[width=0.45\textwidth]{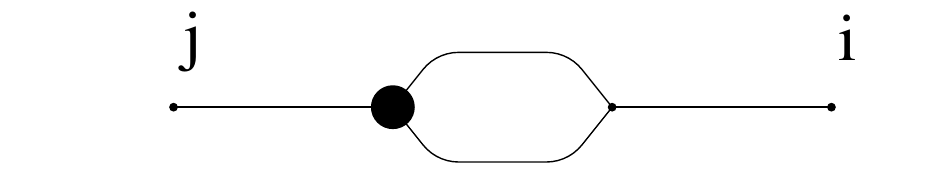}
  \includegraphics[width=0.35\textwidth]{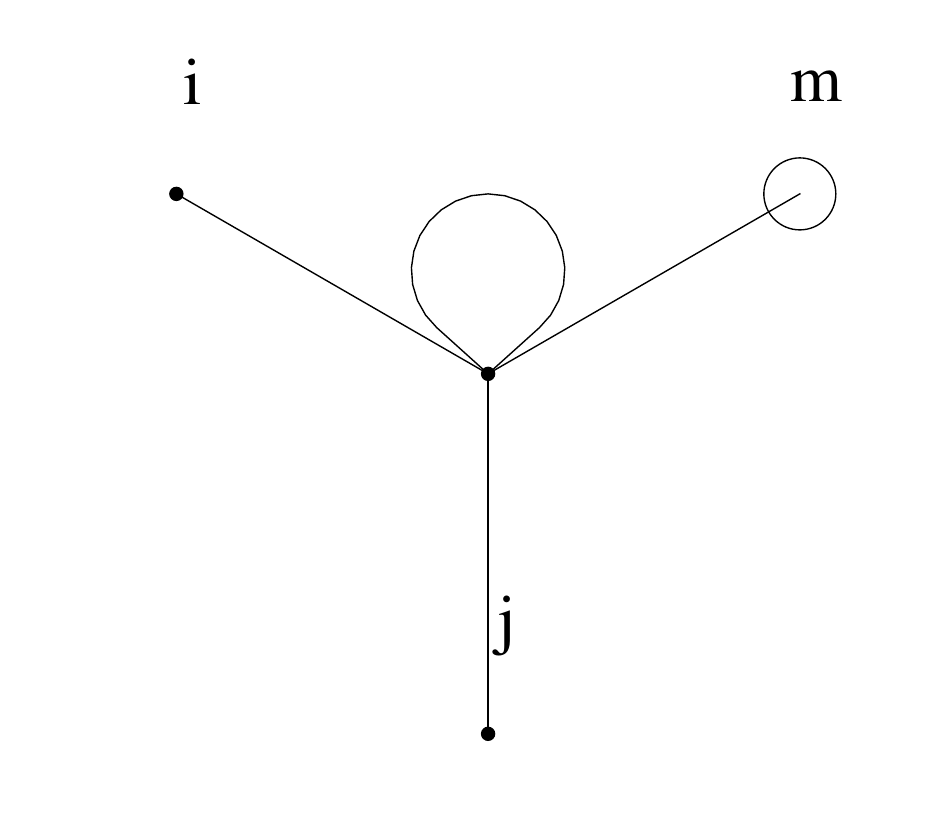}
  \caption{\label{fig:diagrams}Two (simple) examples for loop diagrams as they appear in Dyson-Schwinger equations, 
  see text for details. Solid lines represent a propagator, a blob a dressed vertex and the circle an external field.}
 \end{center}
\end{figure}

\begin{itemize}
 \item Fields are represented by a simple list, e.g. \verb|{A,i}|, where the first entry is the label of the field
and the second entry its generic index. At this level we only use one index which comprises all 
physical indices including momenta. Fields and indices represent the two smallest bits of information. 
 \item In \textit{DoFun} propagators and vertices only have fields as arguments. However, both propagators and fields are allowed as arguments
of the \tw{op} construct. A propagator is defined as \verb|P[{A,a},{B,b}]| where the lists \verb|{A,a}| and \verb|{B,b}| denote its fields.
 \item In the context of functional RG equations a regulator insertion is represented 
by \tw{dR} with two fields as argument, e.~g. \verb|dR[{A,a},{B,b}]|.
 \item In the framework of \textit{DoFun} we have to deal with bare and dressed 
$n$-point functions denoted by \tw{S} and \tw{V}, respectively. 
The arguments of~\tw{S} and~\tw{V} denote the external legs of the vertices. 
For RGEs only dressed $n$-point functions \tw{V} appear.
\end{itemize}
While for bosons the order of fields is irrelevant, it is crucial for fermions. 
We have summarized our conventions for fermions in \ref{app:fermions}. 
With these ingredients it is possible to draw Feynman diagrams with \DoFun.
For example:

\begin{verbatim}
op[S[{W,i},{W,k},{W,l}], P[{W,k},{W,ks}], P[{W,l},{W,ls}], 
 V[{W,ks},{W,ls},{W,j}]]
\end{verbatim}
see left panel of \fref{fig:diagrams}. Here \texttt{W} is a real bosonic field and the external indices of the 
diagram are \texttt{i} and \tw{j}. A tadpole diagram with an additional external leg is given by
\begin{verbatim}
op[S[{W,i},{W,k},{W,l},{W,m},{W,j}], P[{W,k},{W,l}], {W,m}]
\end{verbatim}
see right panel of \fref{fig:diagrams}. Here, the field argument \verb|{W,m}| of the \tw{op} 
function denotes an external field. 
A summary of the basic objects of \textit{DoFun} can be found in Table~\ref{tab:quantDoDSERGE}. 

\begin{table}[tb]
\begin{center}
\begin{tabular}{|l|l|}
 \hline
 Quantity &  Meaning \\ \hline\hline
 \verb|{A,i}| & field \tw{A} with index \tw{i} \\ \hline
 \verb|P[{A,i}, {A,j}]| & propagator of field \tw{A} from \tw{i} to \tw{j} \\ \hline
 \verb|V[{A,i}, ..., {B,j}]| & $n$-point function with legs \verb|{A,i}|, \ldots, \verb|{B,j}|\\ \hline
 \verb|S[{A,i}, ..., {B,j}]| & bare $n$-point function with legs \verb|{A,i}|, \ldots, \verb|{B,j}|\\ \hline
 \verb|dR[{A,i}, {A,j}]| & regulator insertion \\ \hline
 \verb|op[...]| & contains fields, propagators,\\
  & regulator insertions and vertices\\ \hline
\end{tabular}
\caption{\label{tab:quantDoDSERGE} Overview of basic quantities of \textit{DoDSERGE}.}
\end{center}
\end{table}

\begin{table}[tb]
\begin{center}
\begin{tabular}{|l|l|}
 \hline
 Quantity &  Meaning \\ \hline\hline
 \verb|A[mom,i,j,...]| & field \tw{A} with momentum \tw{mom} \\&and indices \tw{i}, \tw{j}, \ldots \\ \hline
 \verb|P[A[mom1,i1,j1,...],A[mom2,i2,j2,...]]| & propagator of field \tw{A} \\ & with indices \tw{i1}, \tw{i2}, \tw{j1}, \tw{j2}, \ldots \\ \hline
 \verb|V[A[mom1,i1,...],...,B[momn,in,...]]| & $n$-point function with legs\\& of fields \tw{A}, \ldots, \tw{B} and \\ & indices \tw{i1}, \ldots, \tw{in} \\ \hline
 \verb|S[A[mom1,i1,...],...,B[momn,in,...]]| & bare $n$-point function with legs\\& of fields \tw{A}, \ldots, \tw{B} and \\ & indices \tw{i1}, \ldots, \tw{in} \\ \hline
 \verb|dR[A[mom1,i1,j1,...],A[mom2,i2,j2,...]]| & regulator insertion \\ \hline
 \verb|op[...]| & contains fields only\\ \hline
\end{tabular}
\caption{\label{tab:quantDoAEFR}Overview of basic quantities of \textit{DoAE} and \textit{DoFR}.}
\end{center}
\end{table}

\subsection{Feynman rules}
\label{ssec:FR}

In order to obtain physically meaningful expressions from the output of \tw{doDSE} and \tw{doRGE} we need to specify the actual 
(physical) meaning of the objects \tw{V}, \tw{P}, \tw{dR} and \tw{S}. Eventually this boils down to a specification of
the Feynman rules. The definition of these objects is left to the
user, i.~e. the user has to provide the details concerning the propagators and the vertices.
The Feynman rules can either be derived by hand or with the aid of independent packages, such as~\textit{FeynRules}~\cite{Christensen:2008py}.
Depending on the package used to derive the Feynman rules, the user may need to adapt the 
sign convention in these rules to ensure that they are compatible with our conventions 
for the definitions of $n$-point functions, see \eref{eq:vertexConvention}.

Alternatively, it is possible to use the package~\textit{DoFR}, which is part of \textit{DoFun}. This package
also allows to derive Feynman rules from a given action.
Independent of the method we choose, we need to define the propagators and vertices to further process the output of the 
commands~\texttt{doDSE} and~\texttt{doRGE}.
In general, this is done by adding appropriate definitions of~\tw{V}, \tw{P}, \tw{dR} and~\tw{S}. 
To be more specific, a propagator in symbolic form is given by
\begin{verbatim}
P[{A,i},{A,j}]
\end{verbatim}
Here, \tw{A} can be any bosonic field. Let us now assume that \tw{A} is a gluon field, i.~e. 
it has Lorentz and color indices. The explicit representation of the propagator is then given by
\begin{verbatim}
P[A[p1, mu, a], A[p2, nu, b]]
\end{verbatim}
The arguments are the fields whose arguments are the momenta and the (color) indices. 
Here, \tw{mu} and \tw{nu} are Lorentz indices and \tw{a} and \tw{b} are adjoint color indices. 
In order to define the propagator we evaluate the following expression:
\begin{verbatim}
P[A[p1_, mu_, a_], A[p2_, nu_, b_], explicit->True]:=
 delta[color,a,b](metric[mu,nu]-p1[mu]p1[nu]/p1^2)/p1^2/Z[p1^2]
\end{verbatim}
At this point several comments are in order:
\begin{itemize}
 \item The option \tw{explicit->True} is required to force \textit{DoFun} to use this 
definition of the propagator, see also below the function \tw{getAE}.
 \item The representation of objects like Kronecker deltas, four-momenta or dressing functions is 
completely left to the user. 
For illustration we have used a self-evident notation here. For example, \tw{metric[mu,nu]} denotes the metric tensor and \tw{p1[mu]} is a four-vector.
 \item The momentum conservation of the propagator is taken into account implicitly, i.~e. 
we define the propagator as a function depending on two different momenta but we only use one of these arguments on the right-hand side. 
Note that the output of~\tw{getAE} already takes momentum conservation into account.
\end{itemize}

Other quantities, such as the vertices \tw{V} and \tw{S} or the regulator insertions \tw{dR}, 
are defined accordingly, see table \ref{tab:quantDoAEFR}. For example, the dressed ghost-gluon vertex can be defined as follows:
\begin{verbatim}
V[A[p1_, mu_, a_], cb[p2_, b_], c[p3_, c_], explicit -> True]:=
 I g structureConstant[a,b,c]
  (ghgDressing1[p1,p2,p3] p2[mu]+ghgDressing2[p1,p2,p3] p1[mu])
\end{verbatim}

Next, we turn to the package \textit{DoFR} which can be used to derive the Feynman rules. 
The use of the commands of this package requires a physical action expressed in terms of the 
\tw{op}~operator. Note that in this case the syntax of the \tw{op} operator 
differs from the syntax which is used to define a symbolic action:
Only fields are permitted here as an argument for the \tw{op} operator. Moreover they have to 
be given in an explicit form: 
\begin{verbatim}
field[momentum, index1, index2, ...]
\end{verbatim}
where \tw{field} is the name of the field. Its first argument is {\it always} the momentum and the others correspond 
to the labels of the indices of the fields, e.~g. color or Dirac indices. 
The order of the indices is determined by the user. 
Note that the physical action always has to be given in momentum space. In general it is automatically 
assumed that there is an integral over reappearing momentum labels and a sum over reappearing indices. 
Integrals and sums are never written out explicitly in \textit{DoFun}. As an example we consider the action of 
an $N$-component scalar theory 
up to fourth order. First, we need to define \tw{phi} as a field. In addition, we have to
specify its indices. This is done with the command \tw{defineFieldsSpecific}, see Section \ref{ssec:representation} for more details:
\begin{verbatim}
defineFieldsSpecific[{phi[momentum, type]}];
\end{verbatim}
\textit{DoFun} now knows that \tw{phi} is a bosonic field with one index \tw{type}. We can check 
our specifications with
\begin{verbatim}
{bosonQ@phi, Head@phi}

--> {True, boson}
\end{verbatim}
and
\begin{verbatim}
indices[phi]

--> {type}
\end{verbatim}
The action in momentum space,
\begin{align}
\int& \ddotp{q} \phi^i (q)q^2\phi^i (-q)\nnnl
 &+\int \ddotp{q}\ddotp{r}\ddotp{s}\lambda(q,r,s,-q-r-s)\phi^i (q)\phi^i (r)\phi^j(s)\phi^j(-q-r-s),
\end{align}
is then given by
\begin{verbatim}
action=q^2 op[phi[q,i],phi[-q,i]]/2
 +lambda[q,r,s,-q-r-s]/8 op[phi[q,i],
  phi[r,i],phi[s,j],phi[-q-r-s,j]];
\end{verbatim}

To define an action within \textit{DoFun} we have to obey a few rules. For example, the 
employed dummy indices must be unique variables and the momenta must be recognizable
as momenta. 
To facilitate this task we may use the command \texttt{convertAction} which attempts to rewrite an action 
given by the user into such a form:
\begin{verbatim}
actionC=convertAction[action]

--> 1/2 q$744^2 op[phi[q$744, dummy[1]], phi[-q$744, dummy[1]]] + 
 1/8 lambda[q$748, q$749, q$750, -q$748 - q$749 - q$750] 
  op[phi[q$748, dummy[2]], phi[q$749, dummy[2]], 
   phi[q$750, dummy[3]], phi[-q$748 - q$749 - q$750, dummy[3]]]
\end{verbatim}
The employed momenta are uniquely labeled \tw{q\$i}. The internal indices in \tw{action} have been replaced by the placeholders~\tw{dummy[...]}
which represent unique dummy indices.

Finally we can use the command \tw{getFR} to derive the two-point and four-point functions:
\begin{verbatim}
twoPoint=getFR[actionC, {phi[p1, i], phi[p2, j]}]

--> p1^2 delta[type, i, j] deltam[p1 + p2]
\end{verbatim}
Here, the $\delta$ distribution in momentum space is represented by 
\begin{verbatim}
deltam[p1+p2]
\end{verbatim}
which corresponds to $(2\pi)^d \de(p1+p2)$
and \tw{delta[type, i, j]} denotes a standard Kronecker delta with indices \tw{i} and \tw{j}. 
We shall return to the function \tw{delta} below.

The general syntax of \tw{getFR} is
\begin{verbatim}
getFR[actionC, {field1[momenta and indices], 
 fields2[momenta and indices], ...}]
\end{verbatim}
From this we infer how to get the four-point function:
\begin{verbatim}
fourPoint=getFR[actionC,
 {phi[p1, i], phi[p2, j], phi[p3, k], phi[p4, l]}];
\end{verbatim}
In the present example the four-point functions consists of 24~terms.
This is the most general case but usually an approximation suffices.
For instance, for a study of an $O(N)$ model a point-like approximation to the quartic coupling
may already be sufficient to capture a wide range of physics.\footnote{In an RG approach the quartic coupling
is a scale-dependent quantity, even in the point-like limit.}
In this case we are left with three terms:
\begin{verbatim}
fourPoint/.lambda[___]:>lambda

--> -lambda delta[type, i, l] delta[type, j, k] 
  deltam[p1 + p2 + p3 + p4] - 
 lambda delta[type, i, k] delta[type, j, l]
  deltam[p1 + p2 + p3 + p4] - 
 lambda delta[type, i, j] delta[type, k, l]
  deltam[p1 + p2 + p3 + p4]
\end{verbatim}
The minus sign is due to our definition of the vertices, see \eref{eq:vertexConvention}.

Having derived the equations for the two- and four-point functions we now have to define the 
specific form of the propagator and the vertex functions \tw{P} and \verb|V|, respectively. 
We infer the form of the propagator from \tw{twoPoint}, which has been calculated above:
\begin{verbatim}
P[phi[p1_,ind1_], phi[p2_,ind2_], explicit->True]:=
 delta[type,ind1,ind2]/p1^2/Z[p1^2];
\end{verbatim}
where we added the dressing function since \verb|P| represents the full propagator.
It is important to stress that no $\delta$ distribution should appear in the definition
of the propagator. The latter is implicitly assumed.
Similar to the propagator we can define the vertex. While the propagator is defined 
to be the inverse of the two-point function, we can directly employ the output of \tw{getFR} to define
the vertex function:
\begin{verbatim}
V[phi[p1_, i_], phi[p2_, j_], phi[p3_, k_], phi[p4_, l_],
 explicit -> True]=
  getFR[actionC, 
   {phi[p1, i], phi[p2, j], phi[p3, k], phi[p4, l]}]/
  deltam[p1 + p2 + p3 + p4] // Expand;
\end{verbatim}
We would like to stress again that no momentum-conserving $\delta$ function should appear in the definition of the vertex functions. Bare vertices \tw{S} and regulator insertions \tw{dR} are defined in the same way as the 
vertex functions.

Finally we would like to comment on a function offered by \textit{DoFun} which may prove useful 
in many cases, namely \tw{delta}. 
It is a generalization of the standard Kronecker delta which allows to relate the indices to 
a certain type of object. For example, \tw{delta[ind, i, j]} is a Kronecker delta with indices \tw{i} and \tw{j}
associated with the index type \tw{ind}. Let us give a few basic examples for 
the application of the function \verb|delta|:
\begin{verbatim}
delta[1,1] --> 1
delta[0,3] --> 0
delta[ind,a,a] --> dim[ind]
\end{verbatim}
where \verb|dim| represents the dimension of the representation associated with this index. 
Although it is not mandatory to specify the type of object to which the indices belong, we highly recommend
to do so in more involved cases with several different indices and types of objects. 
With the function \verb|integrateDeltas| \DoFun\ offers the possibility to 'integrate out' these \tw{delta} 
functions. Usually this function even works in cases in which one of the indices
is part of a different function, e.~g.
\begin{verbatim}
someFunction[a, b] delta[a, c] // integrateDeltas

--> someFunction[c, b]
\end{verbatim}
Note that \verb|integrateDeltas| checks whether an index appears more than twice in \tw{delta} functions.
The functions \tw{delta} and \tw{integrateDeltas} are also used by \tw{getFR}. 
In the subsequent sections we discuss the usage of the main functions of \DoFun\ by means of 
specific examples.

\section{Derivation of functional equations}
\label{sec:programDetails}

\subsection{Derivation of Dyson-Schwinger equations}
\label{ssec:derDSEs}

\begin{figure}[tb]
 \begin{center}
  \includegraphics[width=\textwidth]{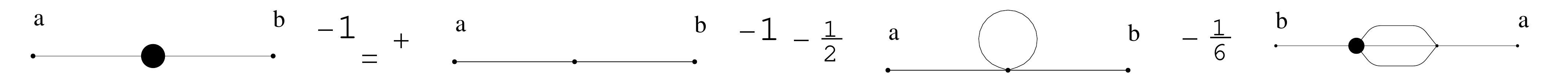}
  \caption{\label{fig:scalar2p}The two-point function DSE of $\vp^4$ theory.}
 \end{center}
\end{figure}

The function for the derivation of DSEs is \tw{doDSE}. In the simplest case the function call reads
\begin{verbatim}
doDSE[action, derivatives]
\end{verbatim}
The main input is the symbolic action which is called \tw{action} in the example above\footnote{In principle, \tw{doDSE} and \tw{doRGE} 
also accept physical actions. However, the physical actions are transformed into symbolic actions by~\tw{doDSE} and \tw{doRGE} .}, 
see Section~\ref{ssec:actions} for details. The argument~\tw{derivatives} is a list of fields corresponding to the legs of the $n$-point function we would like
to compute. A simple example is the DSE for the two-point function of a scalar theory, see also \fref{fig:scalar2p}:
\begin{verbatim}
doDSE[{{phi,phi}, {phi,phi,phi,phi}}, {phi, phi}]

--> op[S[{phi, i1}, {phi, i2}]] - 
 1/2 op[S[{phi, i1}, {phi, i2}, {phi, r1}, {phi, s1}], 
  P[{phi, r1}, {phi, s1}]] - 
 1/6 op[S[{phi, i1}, {phi, r1}, {phi, r2}, {phi, s1}], 
  P[{phi, r1}, {phi, s2}], P[{phi, r2}, {phi, t2}], 
  P[{phi, s1}, {phi, u2}], 
  V[{phi, i2}, {phi, s2}, {phi, t2}, {phi, u2}]]
\end{verbatim}
When we do not attach indices to the fields in the list~\tw{derivatives}, the 
indices are attached automatically to the legs by \tw{doDSE}. In the present example, \tw{doDSE} has chosen \tw{i1} and \tw{i2} as index labels. 
Alternatively, it is possible to choose specific indices by including them in the list~\tw{derivatives}:
\begin{verbatim}
twoPointPhi=doDSE[{{phi,phi}, {phi,phi,phi,phi}}, {{phi,a}, {phi,b}}]

--> op[S[{phi, a}, {phi, b}]] - 
 1/2 op[S[{phi, a}, {phi, b}, {phi, r1}, {phi, s1}], 
  P[{phi, r1}, {phi, s1}]] - 
 1/6 op[S[{phi, a}, {phi, r1}, {phi, r2}, {phi, s1}], 
  P[{phi, r1}, {phi, s2}], P[{phi, r2}, {phi, t2}], 
  P[{phi, s1}, {phi, u2}], 
  V[{phi, b}, {phi, s2}, {phi, t2}, {phi, u2}]]
\end{verbatim}

\begin{figure}[tb]
 \begin{center}
  \includegraphics[width=\textwidth]{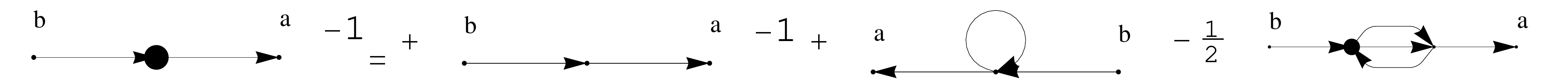}
  \caption{\label{fig:fermion2p}The two-point function DSE of $(\bar{\psi}\psi)^2$ theory.}
 \end{center}
\end{figure}

\begin{figure}[tb]
 \begin{center}
  \includegraphics[width=\textwidth]{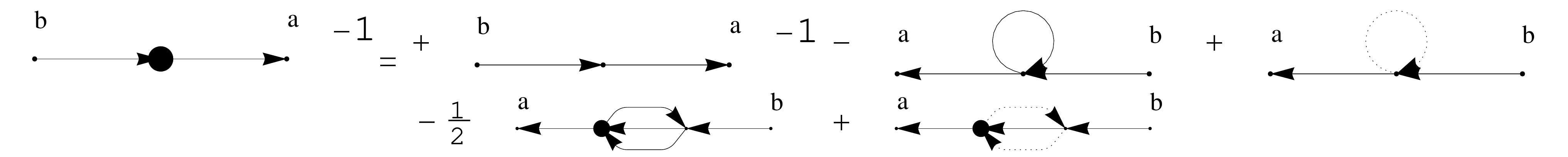}
  \caption{\label{fig:complex2p}The two-point function DSE of a theory with complex scalars $\vp$ and fermions $\psi$ 
  with interactions $(\bar{\vp}\vp)^2$, $(\bar{\psi}\psi)^2$ and $\bar{\vp}\vp\bar{\psi}\psi$. Bosons are denoted by a continuous line, fermions by a dotted one.}
 \end{center}
\end{figure}

For a fermionic theory the derivation of functional equations works in the same way. Consider
\begin{verbatim}
twoPointPsi=doDSE[{{psi, psib}, {psib, psib, psi, psi}}, 
 {{psib, a}, {psi, b}}]

--> op[S[{psi, b}, {psib, a}]] + 
 op[S[{psib, a}, {psib, r1}, {psi, s1}, {psi, b}], 
  P[{psi, s1}, {psib, r1}]] - 
 1/2 op[S[{psib, a}, {psib, r1}, {psi, r2}, {psi, s1}], 
  P[{psi, s2}, {psib, r1}], P[{psi, r2}, {psib, t2}], 
  P[{psi, s1}, {psib, u2}], 
  V[{psib, t2}, {psib, u2}, {psi, s2}, {psi, b}]]
\end{verbatim}
Here, \tw{psi} is the fermion and \tw{psib} the anti-fermion. The first entry in the list \tw{action} corresponds to 
the kinetic term whereas the second term corresponds to a four-fermion interaction, see also Sect.~\ref{sec:GN}.
The result is depicted in terms of Feynman diagrams in~\fref{fig:fermion2p}.
The notation for fermions as \verb|{fermion, anti-fermion}| is quite convenient and is used throughout \DoFun. 
However, the treatment of complex bosonic fields also requires different labels for the fields. To avoid the identification of complex
bosonic fields as Grassmann fields we have to use the option~\tw{specificFieldDefinitions} in \tw{doDSE}.
To be specific, we consider an action which describes a toy model of a complex scalar field \tw{phi} and a fermionic field~\tw{psi}:
\begin{verbatim}
twoPointMixed=doDSE[{{phi, phib}, {psi, psib},
  {psib, psi, phib, phi}, {phib, phib, phi, phi},
  {psib, psib, psi, psi}}, 
 {{phi, b}, {phib, a}}, 
 specificFieldDefinitions -> {phi, phib, {psi, psib}}];
\end{verbatim}
The arguments of \tw{specificFieldDefinitions} are the bosonic fields and a list of the fermion pair.
The output is shown graphically in \fref{fig:complex2p}.

\subsection{Derivation of renormalization group equations}
\label{ssec:doRGE}

\begin{figure}[tb]
 \begin{center}
  \includegraphics[width=0.3\textwidth]{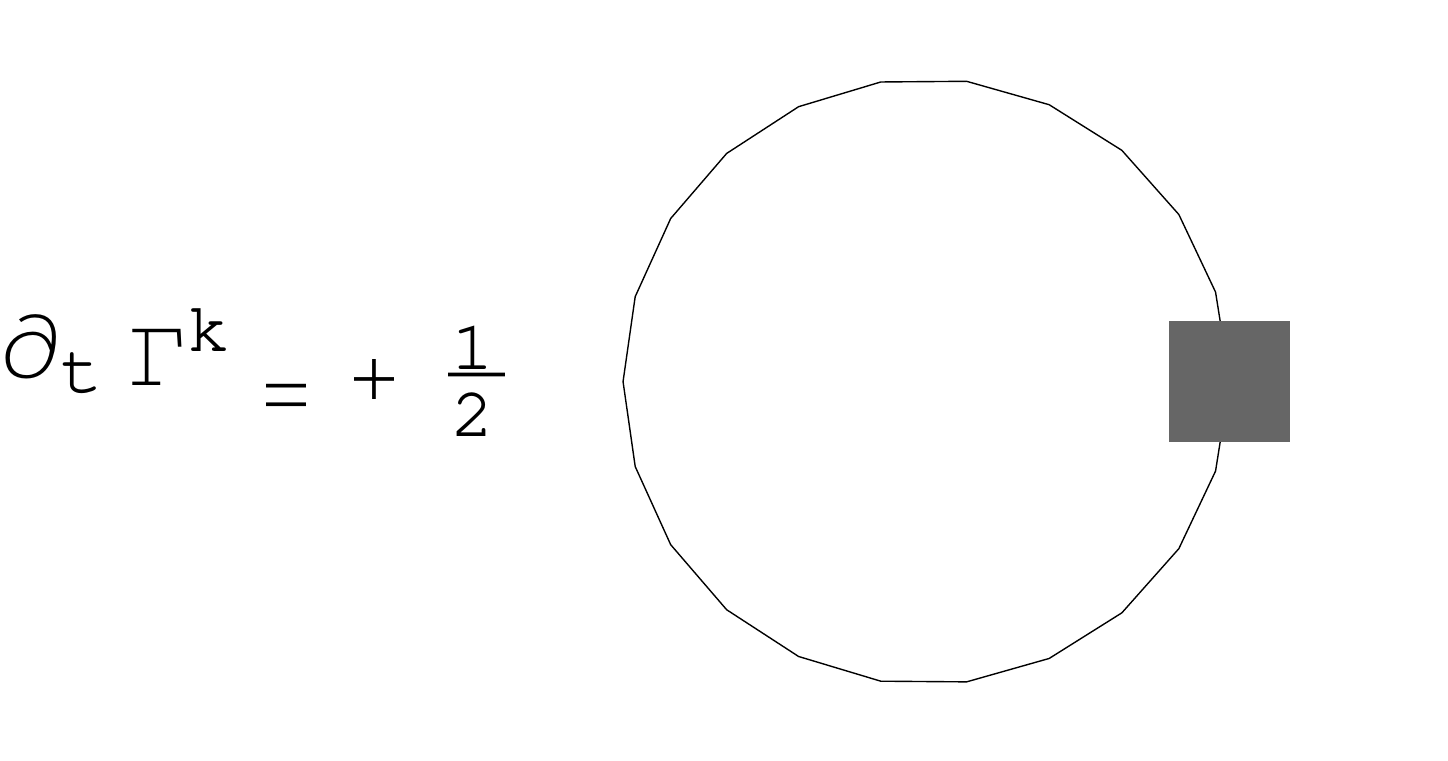}\hfill
  \includegraphics[width=0.6\textwidth]{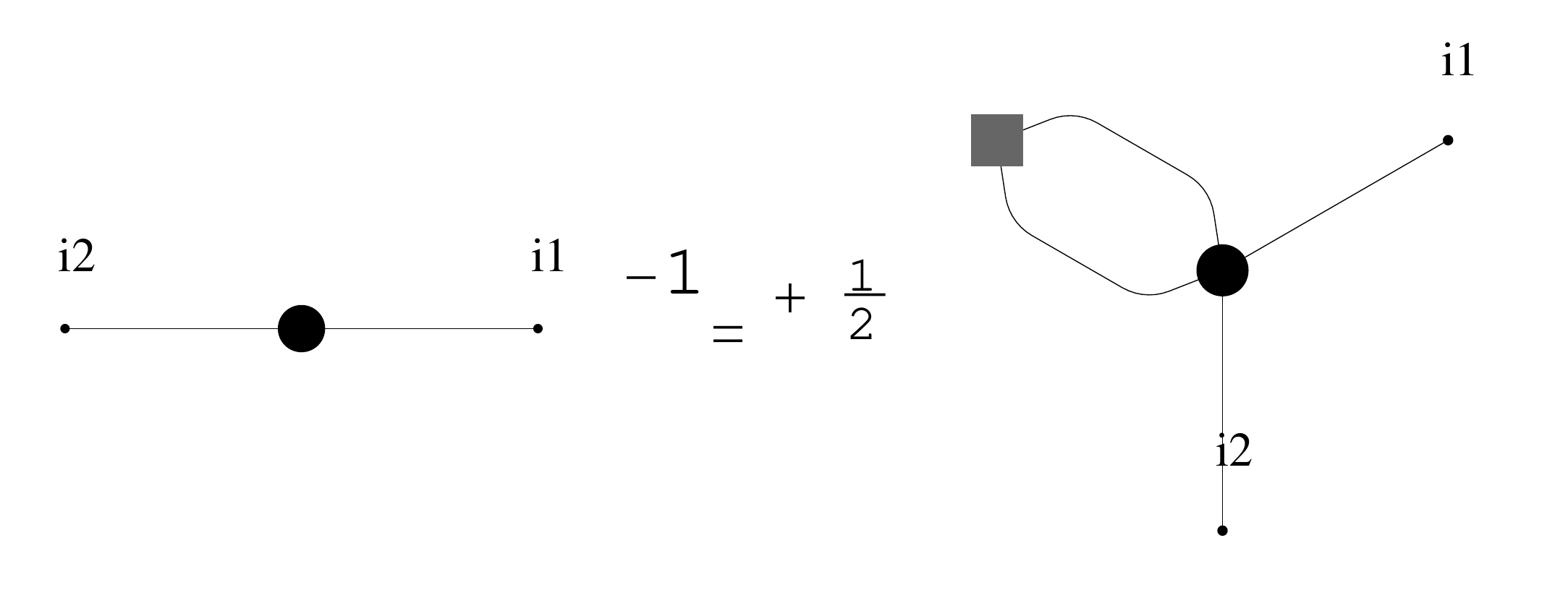}
  \caption{\label{fig:scalarRGE0a2p}The flow equations for the effective action (left) and the two-point function (right).}
 \end{center}
\end{figure}

The derivation of RGEs with the command \tw{doRGE} works very similar to the derivation of DSEs. 
We start with two simple examples: the flow equations for the effective average action itself 
and the two-point function for a $\varphi^4$-theory, see \fref{fig:scalarRGE0a2p}:
\begin{verbatim}
doRGE[{{phi, phi}, {phi, phi, phi, phi}}, {}]

--> 1/2 op[dR[{phi, r1}, {phi, s1}], P[{phi, s1}, {phi, r1}]]
\end{verbatim}
and
\begin{verbatim}
doRGE[{{phi, phi}, {phi, phi, phi, phi}}, {phi, phi}]

--> 1/2 op[dR[{phi, r1}, {phi, s1}], P[{phi, t1}, {phi, r1}], 
  P[{phi, s1}, {phi, v1}], 
  V[{phi, i2}, {phi, i1}, {phi, v1}, {phi, t1}]]
\end{verbatim}
The rules for fermions and complex fields are the same as for the derivation of DSEs.

\begin{figure}[tb]
 \begin{center}
  \includegraphics[width=0.85\textwidth]{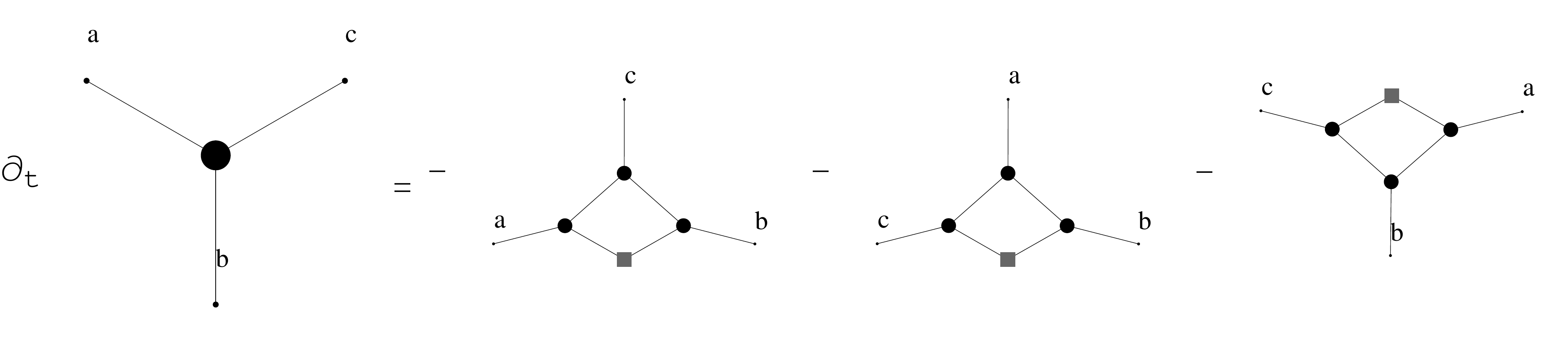}\\
  \includegraphics[width=0.43\textwidth]{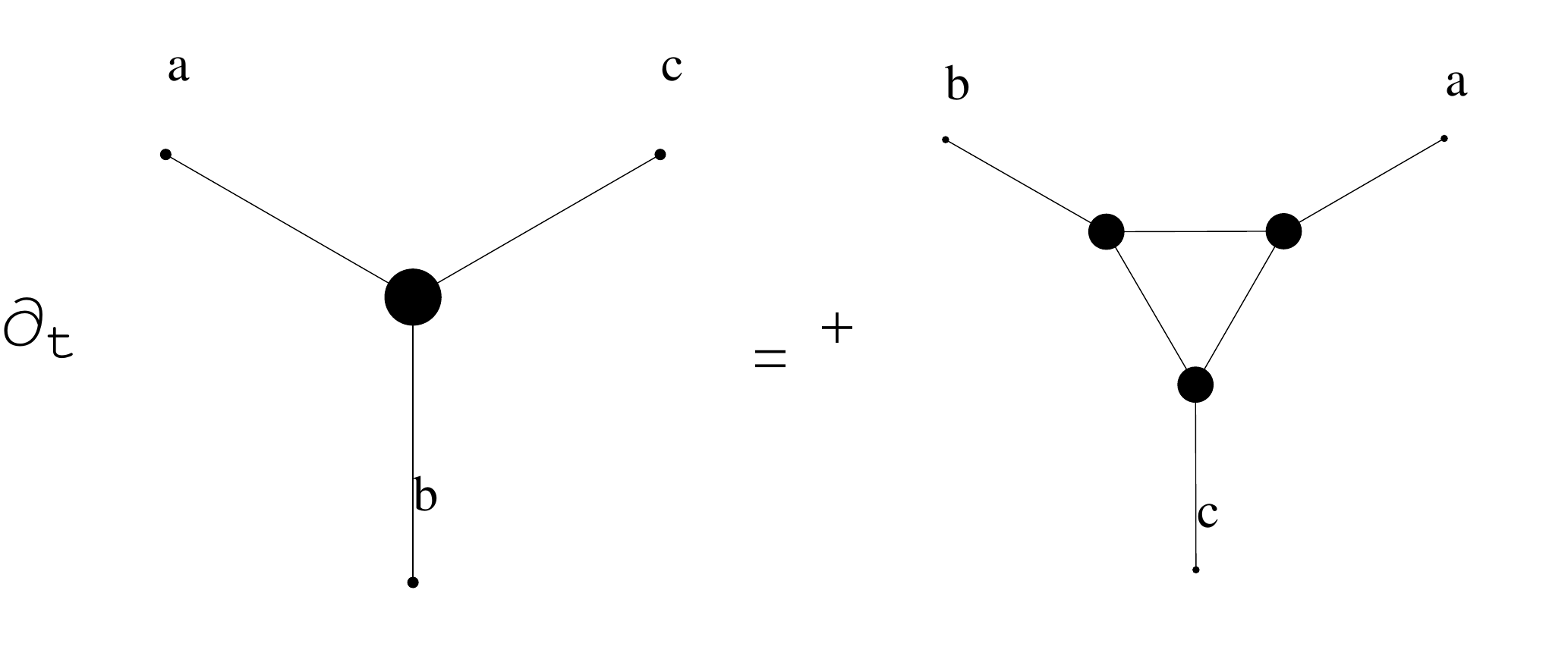}
  \caption{\label{fig:tDeriv}Output of \tw{doRGE} for the three-point function of a $\vp^3$ theory in terms of Feynman diagrams 
  as obtained for \tw{tDerivative->True} (top panel) and for \tw{tDerivative->False} (bottom panel). Note that the derivative $\partial_{\tilde{t}}$ is not indicated explicitly on the right-hand side of the bottom panel though it is contained in the expression.}
 \end{center}
\end{figure}

For~\tw{doRGE} we have the additional option~\tw{tDerivative}. It allows to suppress the derivative $\tilde{\partial}_t$ 
which attaches the derivative of the regulator to the internal lines of the $1$PI diagrams\footnote{Note that diagrams with the same
topology but the regulator insertion attached to different internal lines appear on the right-hand side of the functional RG equations.}. This 
is useful if one is only interested in the structure of the RG equations. For illustration, we show the output of \tw{doRGE} for the three-point function of 
a $\vp^3$ theory in~\fref{fig:tDeriv} as obtained for \tw{tDerivative->True} (top panel) and for \tw{tDerivative->False} (bottom panel). To obtain the output shown 
in the top panel of~\fref{fig:tDeriv} we have used
\begin{verbatim}
threeR = doRGE[{{phi, phi}, {phi, phi, phi}}, {phi, phi, phi}];
\end{verbatim}
whereas we have employed 
\begin{verbatim}
threeRnot = doRGE[{{phi, phi}, {phi, phi, phi}}, {phi, phi, phi}, 
 tDerivative -> False]

--> op[V[{phi, i1}, {phi, r1}, {phi, r2}], P[{phi, r2}, {phi, s2}], 
V[{phi, i2}, {phi, s2}, {phi, t2}], P[{phi, t2}, {phi, u2}], 
V[{phi, i3}, {phi, u2}, {phi, v2}], P[{phi, r1}, {phi, v2}]]
\end{verbatim}
to obtain the output shown in the bottom panel. Here, we have not given the non-graphical output of \tw{doRGE} for the case with \tw{tDerivative->True} 
since it is rather lengthy. Note the different signs appearing in the two graphical representations 
which arise due to the derivative \eref{eq:dtD}.

\subsection{(Graphical) Representation of the output of \tw{doRGE} and \tw{doDSE}}
\label{ssec:representation}
The output of \tw{doDSE} and \tw{doRGE} in terms of \tw{op} functions does not look very transparent from a 
physical point of view. In the previous sections we have already made use of the fact that the output of \tw{doDSE} and \tw{doRGE} can be illustrated
in terms of Feynman diagrams. Before we discuss this option in more detail, we would like to discuss a further possibility
to represent the output of \tw{doDSE} and \tw{doRGE} within \DoFun.

The function~\tw{shortExpression}, or equivalently \tw{sE}, transforms the fields, propagators, regulator insertions and vertices
in the output of \tw{doDSE} and \tw{doRGE}  into more familiar
expressions. For example, 
a propagator is denoted by $\Delta$. Consider
the expression:
\begin{verbatim}
shortExpression[op[S[{W,i},{W,k},{W,l}], P[{W,k},{W,ks}],
 P[{W,l},{W,ls}], V[{W,ks},{W,ls},{W,j}]]]
\end{verbatim}
This yields 
\begin{align*}
 S^{i\, r1\, s1}_{W\, W\, W} \Gamma^{t1\, v1\, j}_{{W\, W\, W}} \Delta^{r1\, t1}_{W\, W} \Delta^{s1\, v1}_{W\, W}\,.
\end{align*}
The field types are given as subscripts and their (collective) indices (momenta, $\dots$) are given as 
superscripts. Propagators are represented as $\Delta$ and bare and dressed vertices as $S$ and $\Gamma$, 
respectively. A regulator insertion is denoted by $\partial_t R$. Note that these string objects can be changed with the aid of the variables 
given in table \ref{tab:symbols}. Moreover \tw{shortEexpression} accepts style options, e.~g.
\begin{verbatim}
shortExpression[op[S[{W,i},{W,k},{W,l}], P[{W,k},{W,ks}],
  dR[{W,ks},{W,ms}], P[{W,ms},{W,m}], P[{W,l},{W,ls}], 
  V[{W,m},{W,ls},{W,j}]], FontSize->20, Bold]
\end{verbatim}
It yields
\begin{align*}
 \begin{large}\boldsymbol{\partial _tR_{W W}^{s1 t1} S_{W W W}^{i r1 r2} 
 \Gamma _{W W W}^{u1 v1 j} \Delta _{W W}^{r1 s1} \Delta _{W W}^{r2 v1} \Delta _{W W}^{t1 u1}} \end{large}.
\end{align*}

\begin{table}[tb]
 \begin{center}
  \begin{tabular}{|l|c|}
   \hline
    Symbol & Standard value\\ \hline\hline
    \verb|$propagatorSymbol| & $\Delta$\\ \hline
    \verb|$bareVertexSymbol| & $S$ \\ \hline
     \verb|$vertexSymbol| & $\Gamma$ \\ \hline
    \verb|$regulatorInsertionSymbol| & $R$\\\hline
  \end{tabular}
  \caption{\label{tab:symbols}Symbols used in \tw{shortExpression}.}
 \end{center}
\end{table}

\begin{figure}[tb]
 \begin{center}
  \includegraphics[width=0.95\textwidth]{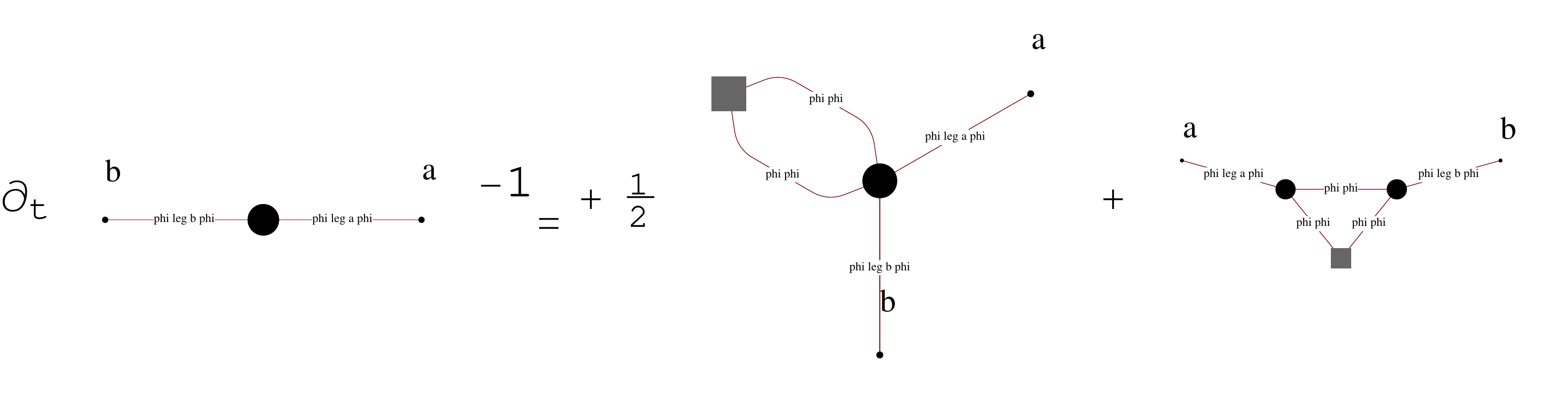}\\
  \includegraphics[width=0.95\textwidth]{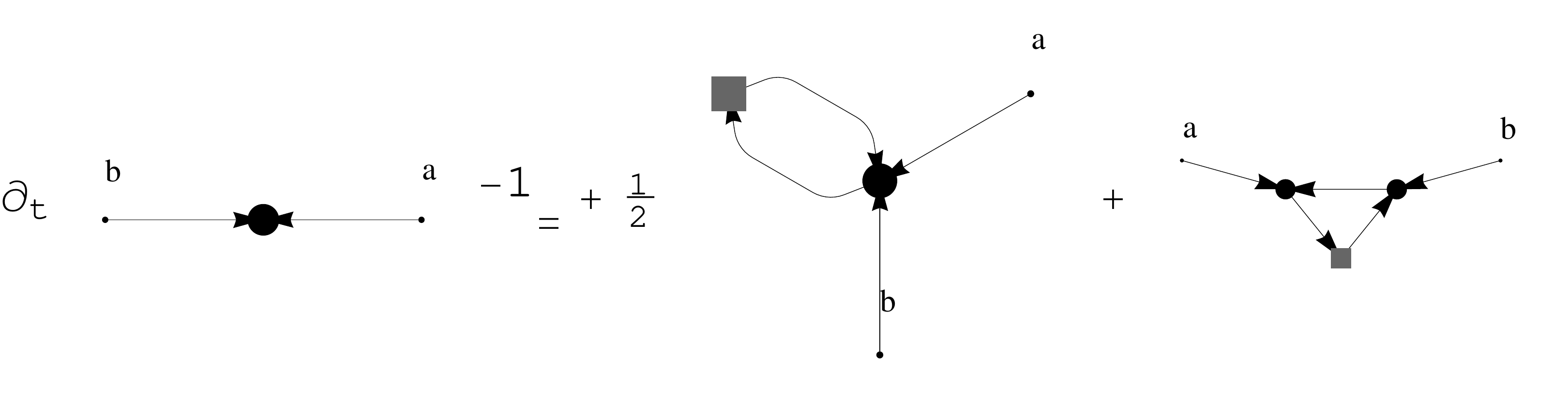}
  \caption{\label{fig:styles}Examples for the representations of RGEs in terms of Feynman diagrams without (top) and with (bottom) user-defined 
  styles for the fields.}
 \end{center}
\end{figure}

Of course, the representation of functional equations in terms of Feynman diagrams is most intuitive from
a physical point of view. To this end, we have included two commands, namely \tw{DSEPlot} and \tw{RGEPlot}. 
As these functions are very similar we only discuss \tw{RGEPlot} here. Differences exist in 
the output, e.~g. \tw{RGEPlot} adds a $\partial_t$ to the left-hand side
of the functional equation. Usually \tw{RGEPlot} is called as follows:
\begin{verbatim}
RGEPlot[output of doRGE, style definitions for the fields]
\end{verbatim}
The first argument is the output of \tw{doRGE}. Of course, modifications by the user are allowed as long as the basic structure, namely a sum of \tw{op} functions, is not changed. 
The second argument is optional. It allows to determine the graphics style for the fields. It has to contain a list of lists: The first argument of each 
sublist gives the name of the field and the remaining arguments can be graphics directives specifying the style of the propagators of that field. 
For the graphics style of the anti-fermions \tw{RGEPlot} automatically uses the same graphics style as for the fermions. 
Style definitions can be, for example, colors or directives like \tw{Dotted} or \tw{Dashed}. If we do not set the style of the fields explicitly, 
the propagators and vertices are labeled according to the names of their fields. Examples for these two cases are
\begin{verbatim}
RGEPlot[
 doRGE[
  {{phi, phi}, {phi, phi, phi}, {phi, phi, phi, phi}},
  {{phi, a}, {phi, b}}]
]
\end{verbatim}
and
\begin{verbatim}
RGEPlot[
 doRGE[
  {{phi, phi}, {phi, phi, phi}, {phi, phi, phi, phi}},
  {{phi, a}, {phi, b}}],
 {{phi, Black}}
]
\end{verbatim}
The output is shown in \fref{fig:styles}.

There are also options which allow to set the style of indices (\tw{indexStyle}) or numerical coefficients (\tw{factorStyle}) and to determine the number of diagrams 
shown in one row. These options are explained in detail in a dedicated part of the \textit{Documentation Center} of \textit{Mathematica}.
Here, we only discuss the option \tw{output}. It allows three settings:
\begin{itemize}
 \item \tw{List}: Gives a list of individual graphs.
 \item \tw{forceEquation}: Draws also the left-hand side of the equation.
 \item \tw{complete} (default): Gives the complete equation including the left-hand side if the expression contains several graphs. If the expression consists of a single graph, only this graph is shown.
\end{itemize}

We would like to point out that we have to define the nature of the fields in order to use \tw{DSEPlot} and \tw{RGEPlot}.
As discussed above, the fields are automatically defined when we use \tw{doDSE} and \tw{doRGE}
but they can also be defined by hand with the aid of the command \tw{defineFields}. Its syntax is
\begin{verbatim}
defineFields[list of bosons, list of fermions,
 list of complex fields]
\end{verbatim}
The lists of fermions and complex fields are given in the usual double notation of fermions and anti-fermions, e.g., \verb|{{c,cb}, {q,qb}}|. If there are no fields of a certain type, an empty list, \verb|{}|, is required as argument.
Note that the nature of the fields is defined automatically when \tw{generateAction} is used.

Both functions, \tw{DSEPlot} and \tw{RGEPlot}, are based on the \Mathematica\ function \tw{GraphPlot}. This has the advantage that we can directly use a 
built-in function of \Mathematica. On the other hand, there are a few drawbacks.
For example, we are not aware of a simple way to plot wiggly lines to, e.~g., represent gluons. Also, for higher vertex functions it 
might be the case that different lines appear on top of each other. Thus, the output looks wrong at first glance. Non-planar diagrams also 
represent a problem for \tw{DSEPlot}, see a corresponding notebook in the \textit{Documentation Center}. 
In general, however, \tw{DSEPlot} and \tw{RGEPlot} provide useful output which
allows to check the associated functional equations.

\subsection{Algebraic expressions}

As illustrative the graphical representation may be, the full analytic expressions containing all indices and momenta are required 
for an evaluation of the functional equations. To this end, we have included the package \textit{DoAE}. 
To translate the output of \tw{doDSE} or \tw{doRGE} into algebraic expressions we use the 
function \tw{getAlgebraicExpression}, or short \tw{getAE}. It uses the symbolic output of \tw{doDSE} or \tw{doRGE} and adds the indices and momenta 
as indicated by the user. Its syntax is
\begin{verbatim}
getAE[exp, external momenta and indices, options]
\end{verbatim}
The first argument is an expression in symbolic notation, not necessarily the output of \tw{doDSE} or \tw{doRGE}. If it is a sum of several \tw{op}-functions, 
\tw{getAE} returns a list so that terms can be traced back to their origin. To obtain the sum instead, one uses the command \tw{Plus@@getAE[...]}. 
The second argument is a list of the external legs together with labels for their momenta and indices. For every external leg there has to be a list entry as follows
\begin{verbatim}
{field, generic index, momentum, real indices}
\end{verbatim}
Here, \tw{generic index} refers to the index label which appears in the output of \tw{doDSE} or \tw{doRGE}. 
Recall that the indices are chosen according to~\tw{i1}, \tw{i2} and so on, 
if we do not explicitly provide index labels. \tw{momentum} can be a symbol, e.~g., \tw{p1}, or a number. 
\tw{real indices} are the indices corresponding to this leg.
Note that Feynman rules have to be properly defined by overloading the functions \tw{P}, \tw{V}, \tw{S} and \tw{dR} 
as described in Section~\ref{ssec:FR}..

The command \tw{getAE} replaces the generic indices appearing in the output of \tw{doDSE} and \tw{doRGE} with 
conventional symbols for the momenta and indices of the fields and
applies the Feynman rules for the propagators and vertices. Therefore we need to define the indices with the function \tw{defineFieldsSpecific}:
\begin{verbatim}
defineFieldsSpecific[{boson[momentum, index1, ...], ...
 {fermion[momentum, index2, ...],
  anti-fermion[momentum, index2, ...]}, ...]
\end{verbatim}
The argument is a list of fields where bosons and fermions are specified in the typical manner of grouping fermion and 
anti-fermion into sublists. The arguments of the fields denote the labels of their indices. \tw{momentum} is the obligatory first argument 
followed by the other indices. The indices of a field can be checked with the command \tw{indices}. Here we do not need to worry about 
complex fields. Note that \tw{defineFieldsSpecific} should not be confused with \tw{defineFields} which is required for plotting graphs.

The names of dummy indices depend on the specific type of index. For example, Lorentz indices, labeled \tw{lor}, are named \verb|\[Mu]|, \verb|\[Nu]| and so on by \tw{getAE}\footnote{Note that the internal routine ensures that these names are unique.}. 
Apart from Lorentz indices color indices in the adjoint representation, labeled \tw{adj}, are the only predefined indices. 
If not specified otherwise, the names \tw{a}, \tw{b}, \tw{c} are assigned automatically to new index labels. 
Alternatively, it is possible to assign labels to specific index types 
with the aid of the function \verb|addIndices|. In fact, this function assigns a list of labels to 
a specific index name, e.~g. flavor:
\begin{verbatim}
addIndices[{flavor, {i, j, k, l, m, n}}]
\end{verbatim}
With \verb|resetIndices[]| we can reset the index labels to the default definitions of \DoFun.

We illustrate the use of \tw{getAE} with the aid of a comparatively simple theory, namely an $N$-component scalar theory truncated at the level of the 
four-point function. Many of the required steps to derive the functional equations of this theory have already been explained above. Therefore we only give the 
commands to derive the flow equations of the two- and four-point functions. First we define the fields
\begin{verbatim}
defineFieldsSpecific[{phi[momentum, type]}];
\end{verbatim}
Next we specify the name of the indices. We choose \tw{i}, \tw{j} and so on:
\begin{verbatim}
addIndices[{type, {i, j, l, m, n}}];
\end{verbatim}
The Feynman rules  for the propagator, the regulator insertion and the quartic vertex with a momentum-independent coupling \tw{lambda}
are defined as follows:
\begin{verbatim}
P[phi[p1_, i_], phi[p2_, j_], explicit -> True] := 
 delta[type, i, j]/(Z[k, p1^2] p1^2 + R[k, p1^2])
dR[phi[p1_, i_], phi[p2_, j_], explicit -> True] := 
 delta[type, i, j] dR[k, p1^2]
V[phi[p1_, i_], phi[p2_, j_], phi[p3_, k_], phi[p4_, l_], 
  explicit -> True] :=
 -lambda delta[type, i, l] delta[type, j, k] - 
  lambda delta[type, i, k] delta[type, j, l] - 
  lambda delta[type, i, j] delta[type, k, l]
\end{verbatim}
To derive the RG equations for the two-point and the four-point functions we use
\begin{verbatim}
twoR = doRGE[{{phi, phi}, {phi, phi, phi, phi}},
  {{phi, a}, {phi, b}}];
fourR = doRGE[{{phi, phi}, {phi, phi, phi, phi}}, 
  {{phi, a}, {phi, b}, {phi, c}, {phi, d}}];
\end{verbatim}
Now we employ \tw{getAE} and choose the external momenta to be zero:
\begin{verbatim}
Plus@@getAE[twoR, {{phi, a, 0, i}, {phi, b, 0, j}}];
Plus@@getAE[fourR, {{phi, a, 0, i}, {phi, b, 0, j}, 
 {phi, c, 0, l}, {phi, d, 0, m}}];
\end{verbatim}
We applied \tw{Plus} to get the sums of the expressions. 
To bring the results into a convenient form we multiply them with appropriate Kronecker $\delta$'s and sum over the corresponding 
indices. For the two-point function we obtain
\begin{verbatim}
integrateDeltas[
  delta[type, j, 1] delta[type, j, 1] Plus @@ 
    getAE[twoR, {{phi, a, 0, i}, {phi, b, 0, j}}]] /. 
 dim[type] :> Ntype

--> -((lambda delta[type, i, j] dR[k, q1^2])/
  (R[k, q1^2] + q1^2 Z[k, q1^2])^2) - 
 (lambda Ntype delta[type, i, j] dR[k, q1^2])/
  (2 (R[k, q1^2] + q1^2 Z[k, q1^2])^2)
\end{verbatim}
For the four-point function we find
\begin{verbatim}
integrateDeltas[
 delta[type, i, 1] delta[type, j, 1] delta[type, l, 1]
 delta[type, m, 1] Plus@@getAE[fourR,
  {{phi, a, 0, i}, {phi, b, 0, j}, {phi, c, 0, l},
   {phi, d, 0, m}}]
 ] /. dim[type] :> Ntype

--> -((24 lambda^2 dR[k, q1^2])/(R[k, q1^2] + q1^2 Z[k, q1^2])^3) -
 (3 lambda^2 Ntype dR[k, q1^2])/(R[k, q1^2] + q1^2 Z[k, q1^2])^3
\end{verbatim}
Specifying the regulator \verb|R[k,q1^2]| and its insertion \verb|dR[k, q1^2]| and adding the integral yields then the final results suitable for an analytic
or numerical evaluation.

This was a first quick introduction into the usage of \tw{getAE}. For a more detailed introduction we refer the reader to the subsequent section. 
There, we also discuss how the equations for a regime with spontaneously broken $O(N)$ symmetry can be derived.

\section{Examples}
\label{sec:examples}

\subsection{O(N) models}\label{sec:ONmodel}

$O(N)$ symmetric scalar field theories play a very prominent role in theoretical physics. First of all, 
they represent a valuable testing ground
for a study of spontaneous symmetry breaking. In particular, the critical exponents at the 
phase transition are accessible to a large variety of methods. Therefore studies of the critical behavior of $O(N)$ models allow us to benchmark theoretical approaches. Within the functional renormalization group approach critical behavior has been indeed studied in great detail, see e.~g. 
Refs.~\cite{Tetradis:1993ts,Litim:2002cf,Braun:2008sg,Benitez:2009xg,Litim:2010tt}.

In this section we use several symbolic actions representing different levels of truncations. First, we define 
an ansatz for an action which contains interactions up to the ten-boson coupling. For the flows calculated below this effectively corresponds to no truncation:
\begin{verbatim}
actionONSymbolic = {{phi, 10, even}};
\end{verbatim}
We then specify a truncated action which only includes a quartic interaction term:
\begin{verbatim}
actionONSTrunc2 = {{phi, 4, even}};
\end{verbatim}
The fields are specified with one index type:
\begin{verbatim}
defineFieldsSpecific[{phi[momentum, type]}];
\end{verbatim}
We define the corresponding physical actions below. However, we would like to add that it is possible to use a potential $U(\phi^2)$  to define the action
 instead of a sum of various interaction terms, e.~g. $U(\phi^2)=(1/2)m^2\phi^2 + (\lambda/8)\phi^4$ where $\phi^{\rm T}=(\phi_1,\dots,\phi_N)$. This 
can be done as follows:
\begin{verbatim}
actionONPotential = convertAction[
 1/2 (q^2 Z[q^2] + R[k]) op[phi[q, a], phi[-q, a]] + U[phi]];
\end{verbatim}
\DoFun\ recognizes \tw{U[phi]} as a self-interaction potential and its derivatives are denoted by \texttt{der[ U[phi], \{phi[p1,i1], ...\}]}.

For the representation of the internal lines in Feynman diagrams we choose 
a solid black line:
\begin{verbatim}
fieldStyleON = {{phi, Black}}; 
\end{verbatim}

\subsubsection{Symmetric regime}

In case of unbroken $O(N)$ symmetry in the ground state we have $\langle  \phi \rangle =0$. We may then use the following 
ansatz for the effective potential and the wave-function renormalization
\begin{align}
U_k(\rho)&=U_k(0)+\sum_{m=1}^\infty \frac{\bar{\lambda}_m}{m!}\rho^m,\\
Z_k(\rho)&=\sum_{m=0}^\infty \frac{Z_k^{(m)}}{m!}\rho^m,
\end{align}
where $\rho=1/2\phi^2$.

\begin{figure}[tb]
 \begin{center}
  \includegraphics[width=0.3\textwidth]{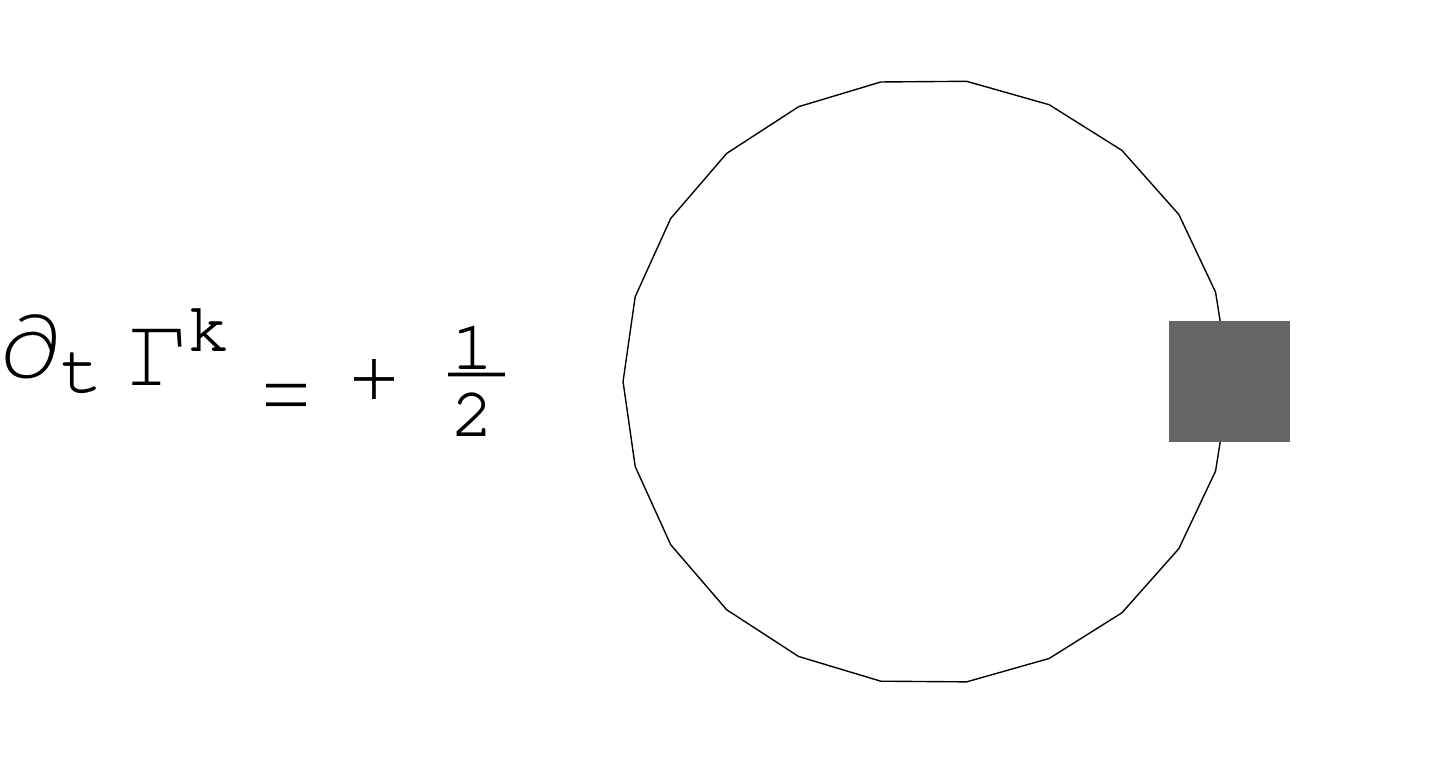}\hfill
  \includegraphics[width=0.6\textwidth]{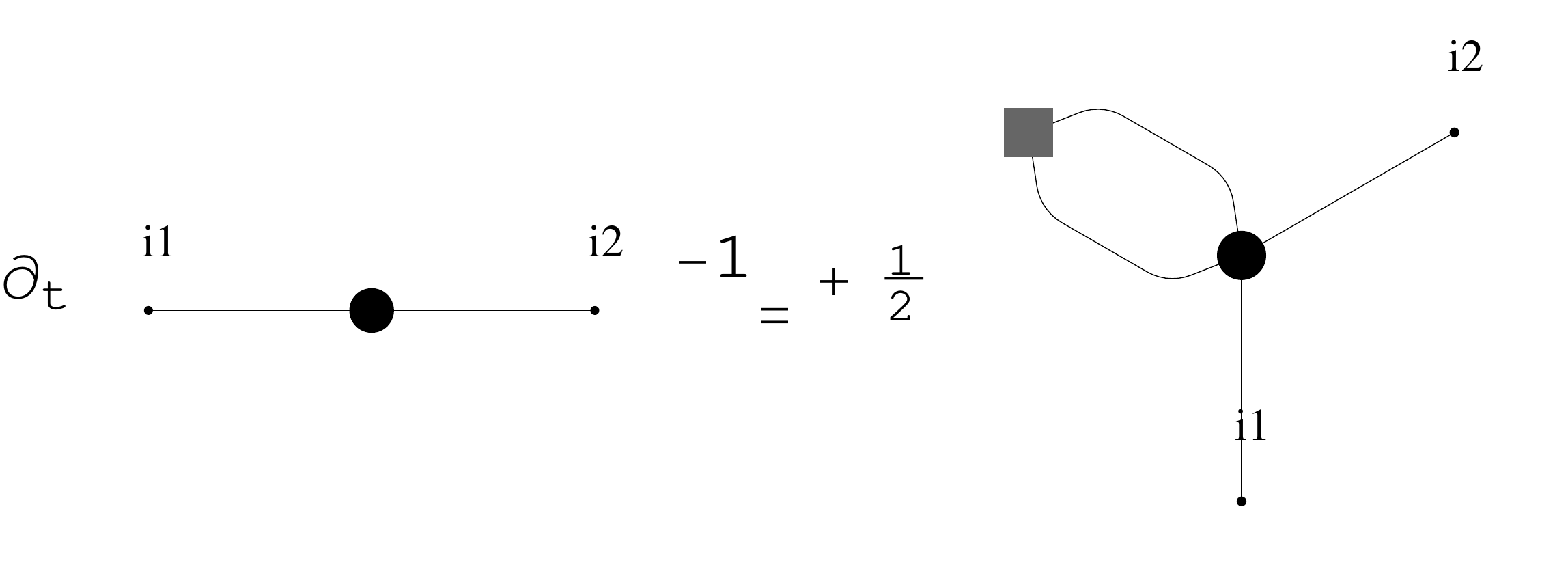}\\
  \includegraphics[width=0.9\textwidth]{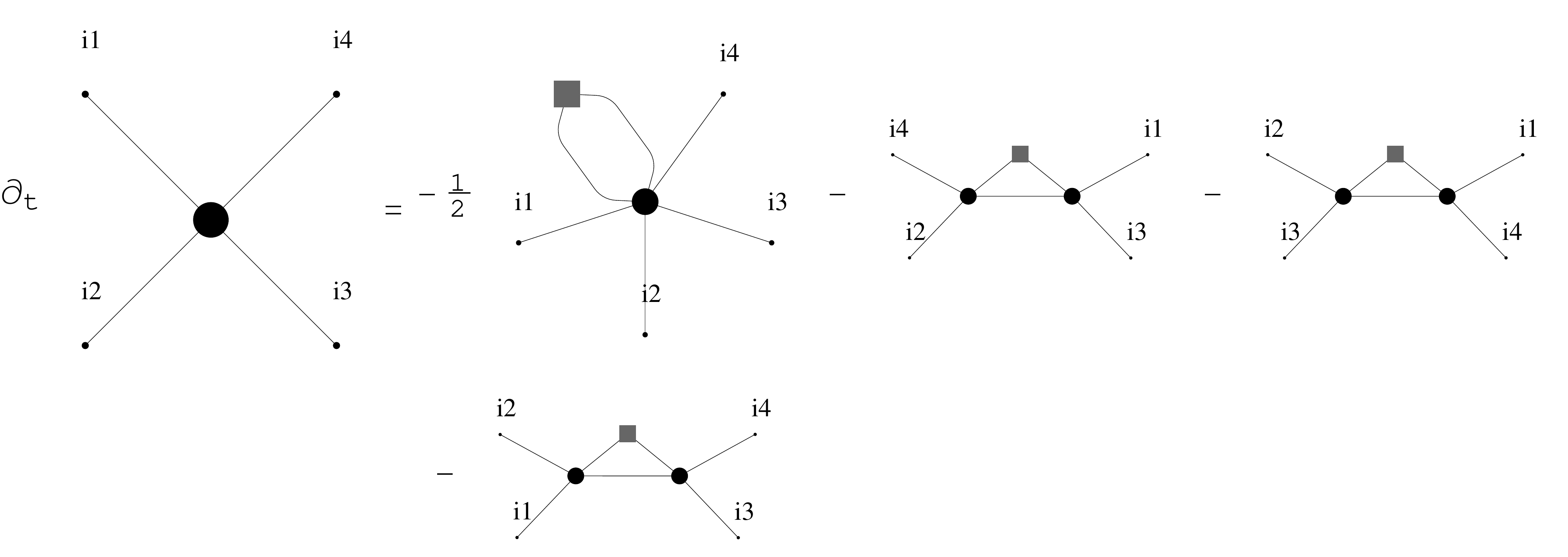}
  \caption{\label{fig:scalarSymmetric}Flow equations for the effective average action and the two- and four-point functions.}
 \end{center}
\end{figure}

The symbolic forms of the flow equations for $U_k(0)$, the two-, four- and six-point functions 
are obtained as follows:
\begin{verbatim}
zeroPoint = doRGE[actionONSymbolic, {}];
twoPoint = doRGE[actionONSymbolic, {phi, phi}];
fourPoint = doRGE[actionONSymbolic, {phi, phi, phi, phi}];
sixPoint = doRGE[actionONSymbolic, {phi, phi, phi, phi, phi, phi}];
\end{verbatim}
The results for the zero-, two- and four-point functions are depicted 
in \fref{fig:scalarSymmetric}. At this level of the derivation these equations are still exact. Higher $n$-point functions 
can also be obtained in this way. However, the computing times increase due to the high number of terms
arising at the intermediate steps of the derivation. For example, it takes roughly two minutes on an AMD Phenom II X4 quad core processor 
to derive the eight-point function; approximately 100.000 terms are generated during the derivation. At the end we are left with only 1.954 distinct diagrams. 
As a non-trivial check we have also derived the flow equation for the nine-point function. This involves about one million intermediate terms. 
As expected due to the underlying symmetries of the theory, these terms sum up to zero.

Let us now define the propagator of the scalar field and the regulator insertion:
\begin{verbatim}
P[phi[p1_, a_], phi[p2_, b_], explicit -> True] := 
 (delta[type, a, b])/(Z[p1^2] + R[p1^2, k]);
dR[phi[p1_, a_], phi[p2_, b_], explicit -> True] := 
  delta[type, a, b] dR[p1^2, k];
\end{verbatim}
In the following we truncate the ansatz for the effective average action at the lowest non-trivial order, i.~e. we only allow
for a quartic interaction term. Thus the physical action reads
\begin{verbatim}
actionON2 = 
  convertAction[
   1/2 (p^2 Z[p^2] + R[k]) op[phi[p, i], phi[-p, i]] + 
    lambda2/8 op[phi[p, i], phi[q, i], phi[r, j], 
      phi[-p - q - r, j]]];
\end{verbatim}
The Feynman rule for the four-point function is then defined as follows:
\begin{verbatim}
V[phi[p1_, i_], phi[p2_, j_], phi[p3_, l_], phi[p4_, m_], 
 explicit -> True] = getFR[actionON2, {phi[p1, i], phi[p2, j],
  phi[p3, l], phi[p4, m]}]/deltam[p1 + p2 + p3 + p4] // Expand;
\end{verbatim}

To obtain the algebraic expressions for the equations we use \tw{getAE}. We start with the flow equation for the 
effective average action:
\begin{verbatim}
zeroPointAlg = Plus @@ getAE[zeroPoint, {}] // integrateDeltas

--> (dim[type] dR[q1^2, k])/(2 (R[q1^2, k] + Z[q1^2]))
\end{verbatim}
To derive the algebraic expressions for the flow equations of the two- and four-point functions with \tw{getAE}
we need to provide lists for the indices and momenta associated with the external legs:
\begin{verbatim}
twoPointAlg = Plus @@ getAE[twoPoint, 
  {{phi, i1, p1, i}, {phi, i2, p2, j}}] // integrateDeltas

--> -((lambda2 delta[type, i, j] dR[(p1 + p2 + q1)^2, 
    k])/((R[(-p1 - p2 - q1)^2, k] + 
     Z[(-p1 - p2 - q1)^2]) (R[q1^2, k] + Z[q1^2]))) -
  (lambda2 delta[type, i, j] dim[type] dR[(p1 + p2 + q1)^2, k])/
    (2 (R[(-p1 - p2 - q1)^2, k] + Z[(-p1 - p2 - q1)^2]) (R[q1^2, k] + 
    Z[q1^2]))
\end{verbatim}
and
\begin{verbatim}
fourPointAlg = Plus @@ getAE[fourPoint, 
  {{phi, i1, 0, i}, {phi, i2, 0, j}, {phi, i3, 0, l},
   {phi, i4, 0, m}}] /. V[a___] :> 0 /; Length@{a} > 4 // 
  integrateDeltas// Simplify

--> -((lambda2^2 (delta[type, i, m] delta[type, j, l] + 
    delta[type, i, l] delta[type, j, m] + 
    delta[type, i, j] delta[type, l, m]) (8 + dim[type]) dR[q1^2, 
   k])/(R[q1^2, k] + Z[q1^2])^3)
\end{verbatim}
Note that we consider the point-like limit here, i.~e. we set the external momenta to zero.
According to our truncation we also set the six-point function to zero with the replacement rule \verb|V[a___] :> 0 /; Length@{a} > 4|. 
Alternatively, we could also define this vertex as 
\begin{verbatim}
V[phi[_], phi[_], phi[_], phi[_], phi[_], phi[_], explicit->True]:=0 
\end{verbatim}
Another way would have been to use the truncated action \tw{actionONSTrunc2} for the calculation of \tw{fourPoint}. The resulting expressions can be evaluated further. This is discussed for the case with broken $O(N)$ symmetry in the following section.

\subsubsection{(Spontaneously) Broken $O(N)$ symmetry in the ground state}
In case of a (spontaneously) broken $O(N)$ symmetry in the ground state 
the expectation value of the field is non-vanishing, $\langle\phi\rangle\neq 0$, and the modes of the theory 
acquire different masses. In fact, we encounter one massive mode and $N\! - \! 1$ Goldstone modes. For 
convenience, we expand the theory about $\langle \phi \rangle =\phi_0 \de^{i1}$ where we consider 
$\phi_0$ to be space-time independent. For simplicity, we shall study a truncation in which we only take into account 
the leading order terms in an expansion of the potential and the wave-function renormalization in powers of~$\rho=(1/2)\phi^2$ :
\begin{align}
U_k(\rho)&= U_k(\rho_0)+\frac{\bar{\lambda}_2}{2}(\rho-\rho_0)^2,\\
Z_k(\rho)&=Z_k.
\end{align}
At finite temperature the constant term in the expansion of the potential $U_k$ is related to thermodynamic observables.
The subscript $0$ denotes the vacuum expectation value, i.e., $\rho_0:=(1/2)\phi_0^2$. The corresponding physical action reads
\begin{verbatim}
defineFieldsSpecific[{phi[momentum, type], phi0[momentum, type]}];
actionON2 = 
 convertAction[
  1/2 (p^2 Z[p^2] + R[k]) op[phi[p, i], phi[-p, i]] + 
   lambda2b/8 op[op[phi[p, i], phi[q, i]] - 
    op[phi0[p, i], phi0[q, i]], 
    op[phi[r, j], phi[-p - q - r, j]] - 
    op[phi0[r, j], phi0[-p - q - r, j]]]];
\end{verbatim}
Note that we have defined \tw{phi0} as a field. This allows us to use it in the derivation of the $n$-point functions.
We have used the \tw{op} function as multiplication operator and therefore \tw{op} functions appear 
as arguments of an \tw{op} function.\footnote{Note that \tw{op[..., op[a], ...]= op[..., a, ...]}.} The Feynman rules are given by

\begin{verbatim}
P[phi[p1_, a_], phi[p2_, b_], explicit -> True] := 
  delta[type, a,  1] delta[type, b, 1]/
   (Z[p1^2] p1^2 + R[p1^2, k] + 2 rho0 lambda2b) + 
  (delta[type, a, b] - delta[type, a, 1] delta[type, b, 1])/
   (Z[p1^2] p1^2 + R[p1^2, k]);

dR[phi[p1_, a_], phi[p2_, b_], explicit -> True] := 
  delta[type, a, b] dR[p1^2, k];

V[phi[p1_, i_], phi[p2_, j_], phi[p3_, k_], phi[p4_, l_], 
  explicit -> True] = 
 getFR[actionON2, {phi[p1, i], phi[p2, j], phi[p3, k], phi[p4, l]}, 
    symmetry -> broken]/deltam[p1 + p2 + p3 + p4] // Expand;
\end{verbatim}
We split the propagator into $N-1$ massless modes (Goldstone modes) and one massive mode. 
For the sake of simplicity, we consider the wave-function renormalization \tw{Z} to be identical 
for the massive and the radial mode. In the derivation of the four-point function we used the option \verb|symmetry -> broken| to indicate that the vacuum expectation value of the field does not vanish.

With these Feynman rules we can employ the symbolic expression for the zero-point function from the previous section to obtain the flow equation for $U_k(\rho_0)$:
\begin{verbatim}
zeroPointAlg = Plus @@ getAE[zeroPoint, {}] // integrateDeltas

--> -(dR[q1^2, k]/(2 (R[q1^2, k] + q1^2 Z[q1^2]))) + 
 (dim[type] dR[q1^2, k])/(2 (R[q1^2, k] + q1^2 Z[q1^2])) + 
 dR[q1^2, k]/(2 (2 lambda2b rho0 + R[q1^2, k] + q1^2 Z[q1^2]))
\end{verbatim}
The left-hand side of the equation reads
\begin{align}
\partial_t \Gamma_k\Big|_{\phi=\phi_0}=
\int d^d x \partial_t U_k(\phi_0)= \partial_t U_k(\phi_0)\int d^d x.
\end{align}
The integral over $d^d x$ corresponds to the momentum conserving $\delta$ distribution on the 
right-hand side, $(2\pi)^d\de(0)$. The latter is not given explicitly in the output of \tw{doRGE} but always implicitly 
assumed. Thus, the flow equation for $U_k(\rho_0)$ reads
\beq
\partial_t U_k(\rho_0)&=&\frac{1}{2}\int \ddotp{q} \Big( (N-1)\frac{\partial_t R_k(q^2)}{Z_k(q^2) q^2+R_k(q^2)}\nonumber\\
&&\qquad\qquad\quad +\frac{\partial_t R_k(q^2)}{Z_k(q^2) q^2+R_k(q^2)+2\bar{\lambda}_2 \rho_0}\Big).
\eeq
This equation agrees with the flow equation found in~Refs.~\cite{Wetterich:1992yh,Tetradis:1993ts}.

For the derivation of the flow equations of $\rho_0$ and the quartic coupling $\bar{\lambda}_2$  
we introduce the renormalized dimensionless mass and coupling:
\begin{align}\label{eq:renormalizedQuants}
\ka&=Z_k k^{2-d}\rho_0\,,\nnnl
\lambda_2&=Z_k ^{-2}k^{d-4} \bar{\lambda}_2\,.
\end{align}

Here, we consider a truncation in which we only include the flow of $\rho_0$ and the 
quartic interaction. Higher-order interactions, such as a six-boson interaction, are set to zero. 
Within \DoFun\ it is convenient to start with the derivation of the flow
equation of the highest-order coupling. This is due to the fact that
the flow equation of the $n$-boson coupling depends directly on the flow of the 
$(n+2)$-boson coupling via a term $\sim \partial_t \phi_0^2$, see Ref.~\cite{Tetradis:1993ts}
and the discussion of the flow of $\rho_0$ below for details.

The left-hand side of the flow equation of the four-point function reads:
\begin{align}\label{eq:ON4plhs}
&\frac{\de^4}{\de \phi^i(x) \de \phi^j(y)\de \phi^l(z) \phi^m(z)}\partial_t \Gamma_k\Big|_{\phi=\phi_0}\nnnl
&\quad= Z_k^2 k^{4-d}(\de^{ij}\de^{lm}+\de^{il}\de^{jm}+\de^{im}\de^{jl})(4-d-2\eta+\partial_t )\lambda_2\de(x-y)\de(x-z)\de(x-u),
\end{align}
where $\eta=-\partial_t \ln Z$. The Fourier transformation for vanishing external momenta yields
\begin{align}
&\frac{\de^4}{\de \phi^i(0) \de \phi^j(0)\de \phi^l(0) \phi^m(0)}\partial_t \Gamma_k\Big|_{\phi=\phi_0}\nnnl
&\quad= Z_k^2 k^{4-d}(\de^{ij}\de^{lm}+\de^{il}\de^{jm}+\de^{im}\de^{jl})(4-d-2\eta+\partial_t )\lambda_2\int d^d x .
\end{align}
On the right-hand side we obtain the algebraic form as follows\footnote{Since we do not take into account
the flow of the six-boson interaction, we do not use the option \tw{symmetry->broken} in \tw{doRGE}. 
At this point this option 
should only be used if one is also interested in a derivation of the flow equation of the six-boson coupling. This is due to the fact that the flow of the quartic interaction then depends directly on the flow of the six-boson coupling,
see Ref.~\cite{Tetradis:1993ts} for details.}:
\begin{verbatim}
fourPointTrunc2 = doRGE[actionONSTrunc2, {phi, phi, phi, phi}];
fourPointT2Alg = 
 Plus @@ getAE[
    fourPointTrunc2, {{phi, i1, 0, i}, {phi, i2, 0, j},
     {phi, i3, 0, l}, {phi, i4, 0, m}}] // integrateDeltas;
\end{verbatim}
where we derived the flow of the four-point function from \tw{actionONSTrunc2} in the first line.
In particular for a study of critical phenomena it is convenient to introduce dimensionless quantities, see \eref{eq:renormalizedQuants}. 
We define the following set of rules to rewrite the output accordingly:
\begin{verbatim}
dimLessRulesON = {R[q1^2, k] :> q1^2 r[y], 
   rho0 :> Z[q1^2]^(-1) kappa k^(d - 2), 
   lambda2b :> Z[q1^2]^2 lambda2 k^(4 - d), 
   dR[q1^2, k] :> q1^2 dtr[y], q1 :> Sqrt[y ] k, dq1 :> k^2/2/q1};
\end{verbatim}
Multiplying \tw{fourPointT2Alg} with $\de^{i1}\de^{j1}\de^{l1}\de^{m1}$ and applying the rules \tw{dimLessRulesON} we find
\begin{verbatim}
angleInt[d_] := 4v[d];
fourPointT2Algy = 
 Simplify[(angleInt[d] dq1 q1^(d - 1)  integrateDeltas[
      delta[type, i, 1] delta[type, j, 1] delta[type, l, 1]
       delta[type, m, 1] fourPointT2Alg] //. dimLessRulesON), 
  d \[Element] Integers]

--> 6 k^(4 - d) lambda2^2 y^(d/2)
  dtr[y] v[d] Z[k^2 y]^4 (1/(y^3 (r[y] + Z[k^2 y])^3) - dim[type]/
   (y^3 (r[y] + Z[k^2 y])^3) - 
    9/(y r[y] + (2 kappa lambda2 + y) Z[k^2 y])^3)
\end{verbatim}
Here we have introduced the integral measure \verb|angelInt[d] dq1 q1^(d - 1)| 
and \tw{v[d]} represents $v_d=(\pi^{d/2} 2^{d + 1} \Gamma(d/2))^{-1}$.
On the left-hand side the projection yields a factor of $3$:
\begin{verbatim}
integrateDeltas[
  delta[type, i, 1] delta[type, j, 1] delta[type, l, 1] delta[type, m,
     1] (delta[type, i, j] delta[type, l, m] + 
     delta[type, i, l] delta[type, j, m] + 
     delta[type, i, m] delta[type, j, l])] //. dimLessRulesON

--> 3
\end{verbatim}
Combining both sides yields
\begin{align}\label{eq:flow-lambda}
\partial_t \lambda_2=&(d-4+2\eta)\lambda_2+2 v_d\lambda_2^2 \int_0^\infty dy \,y^{d/2}\bigg((N-1)\frac{Z_k^2 \partial_t r(y)}{(Z_k+r(y))^3}\nnnl
&\qquad\qquad\qquad\qquad\qquad\qquad\quad +9\frac{Z_k^2 \partial_t r(y)}{(Z_k(y+2\ka \lambda-2)+y r(y))^3} \bigg).
\end{align}
Recall that \tw{doRGE} computes the $n$-point function which is defined to be the negative derivative of the effective average 
action for $n>2$, see \eref{eq:vertexConvention}.

\begin{figure}[tb]
 \begin{center}
  \includegraphics[width=0.9\textwidth]{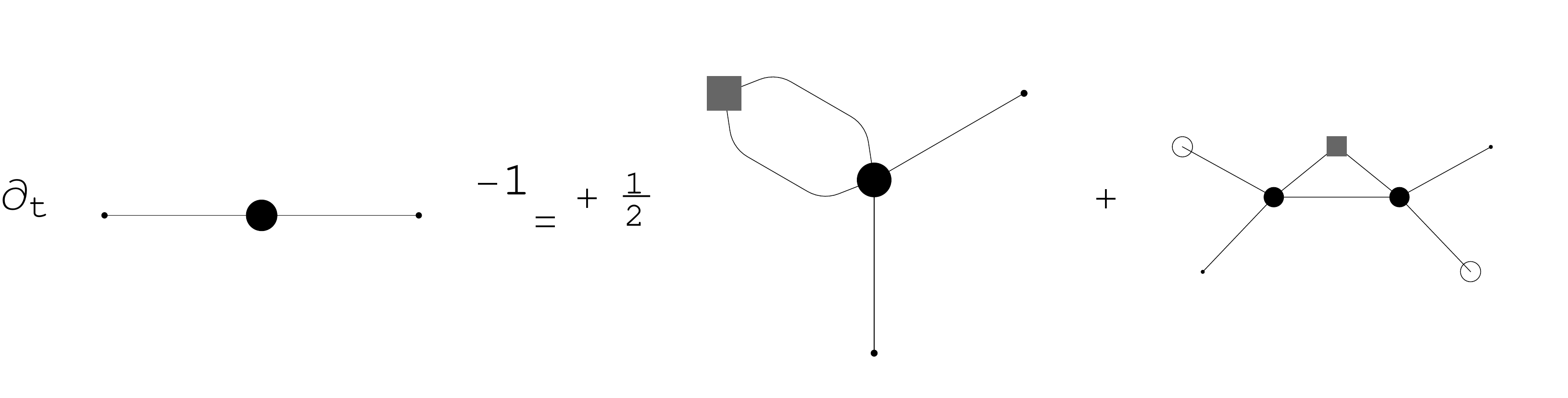}
  \caption{\label{fig:twoPointB} (Truncated) Flow equation for the two-point function in the case of broken $O(N)$ symmetry. The circles correspond to the non-vanishing expectation value of the field.}
 \end{center}
\end{figure}

Next, we derive the algebraic form of the flow equation for the two-point function in the limit of vanishing external momenta. 
To this end, we need to rederive the expression for the two-point function since now additional contributions due to 
the broken $O(N)$ symmetry appear on the right-hand side. The result is depicted in \fref{fig:twoPointB}.
We take care of the broken symmetry in the various steps of the derivation by setting the option \tw{symmetry-> broken} in \tw{doRGE}:
\begin{verbatim}
twoPointB = doRGE[actionONSTrunc2, {phi, phi}, symmetry -> broken];
\end{verbatim}
The output is used as input for \tw{getAE}:
\begin{verbatim}
twoPointAlg = 
  integrateDeltas[
   Plus @@ getAE[twoPointB, {{phi, i1, 0, i}, {phi, i2, 0, j}}]];
\end{verbatim}
The computation of the left-hand side is now slightly more involved than in the case with intact $O(N)$ symmetry. We find
\begin{align}\label{eq:ON2plhs}
&\frac{\de^2}{\de \phi^i(x) \de \phi^j(y)}\partial_t \Gamma_k\Big|_{\phi=\phi_0}\nnnl
&\quad = \frac{\de^2}{\de \phi^i(x) \de \phi^j(y)}\int dz \Big(
 (\partial_t Z_k)\phi^l(z)(-\Box)\phi^l(z)+\frac{\partial_t \bar{\lambda}_2}{2} (\rho-\rho_0)^2\nnnl
&\qquad \qquad\qquad\qquad\qquad\qquad+\bar{\lambda}_2 (\rho-\rho_0)(-\partial_t\rho_0) \Big)\Big|_{\phi=\phi_0}\nnnl
&\quad= \Big(-\de^{ij} \eta Z_k (-\Box)+Z_k^2 k^{4-d}\phi_0^i(x)\phi_0^j(y)(4-d-2\eta+\partial_t )\lambda_2\nnnl
&\qquad  \qquad\qquad\qquad\qquad\qquad+Z_k k^2 \lambda_2 \de^{ij}(2-\eta-d-\partial_t)\ka \Big)\de(x-y).
\end{align}
To project onto the flow of the renormalized mass $\ka$ we multiply both sides with $\de^{i1}\de^{j1}$ and replace the non-vanishing 
fields by $\rho_0$:
\begin{verbatim}
twoPointAlgProj = 
 integrateDeltas[delta[type, i, 1] delta[type, j, 1] twoPointAlg] /. 
   phi[0, i_] :> phi0[i] /. phi0[_]^2 :> 2 rho0

--> -((2 lambda2b^2 rho0 dR[q1^2, k])/(R[q1^2, k] + q1^2 Z[q1^2])^3) +
  (2 lambda2b^2 rho0 dim[type] dR[q1^2, k])/(R[q1^2, k] + 
   q1^2 Z[q1^2])^3 + (lambda2b dR[q1^2, k])/(
 2 (R[q1^2, k] + q1^2 Z[q1^2])^2) - (lambda2b dim[type] dR[q1^2, k])/(
 2 (R[q1^2, k] + q1^2 Z[q1^2])^2) + (
 18 lambda2b^2 rho0 dR[q1^2, k])/(2 lambda2b rho0 + R[q1^2, k] + 
   q1^2 Z[q1^2])^3 - (3 lambda2b dR[q1^2, k])/(
 2 (2 lambda2b rho0 + R[q1^2, k] + q1^2 Z[q1^2])^2)
\end{verbatim}
Let us now apply the rules \tw{dimLessRulesON} to obtain a dimensionless flow equation. This yields
\begin{verbatim}
twoPointAlgProjy = 
 Simplify[#, 
    Element[d, 
     Integers]] & /@ (angleInt[d] dq1 q1^(d - 1) twoPointAlgProj //. 
     dimLessRulesON // Expand)// Simplify

--> k^2 lambda2 y^(d/2) dtr[y] v[d] Z[k^2 y]^2 
 (1/(y^2 (r[y] + Z[k^2 y])^2)
  - dim[type]/(y^2 (r[y] + Z[k^2 y])^2)
  - 3/(y r[y] + (2 kappa lambda2 + y) Z[k^2 y])^2 
  + 4 kappa lambda2 Z[k^2 y] (-(1/(y^3 (r[y] + Z[k^2 y])^3))
   + dim[type]/(y^3 (r[y] + Z[k^2 y])^3) 
   + 9/(y r[y] + (2 kappa lambda2 + y) Z[k^2 y])^3))
\end{verbatim}
On the left-hand side we project with $\de^{i1}\de^{j1}$:
\begin{align}
&\de^{i1}\de^{j1}\frac{\de^2}{\de \phi^i(x) \de \phi^j(y)}\partial_t \Gamma_k\Big|_{\phi=\phi_0}=\nnnl
&\quad= \Big(- \eta Z_k (-\Box)+Z_k k^2 \lambda_2 (2-\eta-d-\partial_t)\ka +\nnnl
&\qquad+2Z_k k^{2}\ka(4-d-2\eta+\partial_t )\lambda_2 \Big)\de(x-y).
\end{align}
In the point-like limit this reduces to 
\begin{align}
&\de^{i1}\de^{j1}\frac{\de^2}{\de \phi^i(0) \de \phi^j(0)}\partial_t \Gamma_k\Big|_{\phi=\phi_0}=\nnnl
&\quad= \left(Z_k k^2 \lambda_2 (2-\eta-d-\partial_t)\ka+2Z_k k^{2}\ka(4-d-2\eta+\partial_t )\lambda_2 \right)  \int d^d x .
\end{align}
Note that we have switched to momentum space to obtain this expression.
The integral over $d^d x$ is associated with the momentum conserving $\delta$ function and cancels against the right-hand side, see also
comments above. 

We observe that terms proportional to $2 k^{2}\ka$ appear on the left-hand side. These terms correspond
to the flow equation of the quartic coupling. Inserting the RG flow equation of the quartic coupling into 
the expression above, we obtain the flow equation for the renormalized and dimensionless vacuum expectation value $\ka$:
\begin{align}
\partial_t \ka &= (2-\eta-d)\ka -v_d \int_0^\infty dy y^{d/2} Z_k\Big( (1-N)\frac{r'(y)}{y^2(r(y)+Z_k)^2} \nnnl
&-\frac{3r'(y)}{(y\,r(y)+(2\kappa \lambda_2+y)Z_k)^2} \Big).
\end{align}

Finally we have to specify a regulator function. To demonstrate how this can be done within \DoFun, we consider the flow of the quartic coupling.
In the following we choose an optimized regulator function~$R_k=(k^2-q^2)\theta(k^2-q^2)$, see~Refs.~\cite{Litim:2000ci,Litim:2001up,Pawlowski:2005xe}.
The momentum integrations can then be performed analytically with the aid of~\textit{Mathematica}. For this purpose, it is convenient to define the rules
\begin{verbatim}
regRulesZ = {r[y] :> (1/y - 1) UnitStep[1 - y], 
   dtr[y] :> 2/y UnitStep[1 - y], Z[_] :> 1};
\end{verbatim}
Applying them to \tw{fourPointT2Algy}, i.~e. to the integral appearing in the flow equation 
of the four-point function calculated above, we find:
\begin{verbatim}
Integrate[
 fourPointT2Algy/(-3 k^(4 - d)) /. regRulesZ, {y, 0, \[Infinity]}, 
 Assumptions -> {d > 0}]

--> -((8 lambda2^2 (1 - 9/(1 + 2 kappa lambda2)^3 - dim[type]) 
 v[d])/d)
\end{verbatim}
In standard notation the complete flow equation reads then
\begin{align}
 \partial_t \lambda_2=(d-2+2\eta)\lambda_2-8\lambda_2^2 \frac{v_d}{d}\left((1-N)-\frac{9}{1+2\ka \lambda_2)^3 }\right).
\end{align}
Note that the same procedure can be applied to the expressions of all flow equations derived above.

\subsection{Gross-Neveu-model}\label{sec:GN}
In the previous section we have discussed a purely bosonic theory. In this section we turn to a purely
fermionic formulation of the Gross-Neveu model.
This model allows to study dynamical chiral symmetry breaking as driven by fermion fluctuations. 
In fact, the finite-temperature phase diagram of the Gross-Neveu model in $d<4$ space-time dimensions has drawn a lot of 
attention in recent years~\cite{Thies:2003kk,Basar:2008im}. Here, we only aim at a study of a purely fermionic formulation
of this theory at vanishing temperature which is sufficient for a first non-trivial check of \DoFun.

The ansatz for the effective action we are going to employ is
\begin{align}\label{eq:ansatzGN}
 S[\bar{\psi},\psi]=& \sum_{j=1}^{N_f} \int \ddotp{q}Z(q)\bar{\psi}_j(q) (-\slashed{q}) \psi(ßq)\nnnl
 &+\sum_{i,j=1}^{N_f}\int \ddotp{q_1}\ddotp{q_2}\ddotp{q_3} \bar{\psi}_i(q_1)\psi_i(q_2)\frac{\bar{g}}{2N_f}\bar{\psi}_j(q_3)\psi_j(-q_1-q_3-q_2).
\end{align}
We first define the fields \tw{psi} and \tw{psib} as fermion and anti-fermion, respectively, with a Dirac and a flavor index,
\begin{verbatim}
defineFieldsSpecific[{{psi[momentum, dirac, flavor], 
 psib[momentum, dirac, flavor]}}];
\end{verbatim}
and then the action:
\begin{verbatim}
actionGN = convertAction[
 diracM[q, a, b] (Z[q^2] + Rs[q^2, k]) op[psib[q, a, i], 
  psi[q, b, i]] + 
 gb/(2 Nf) op[psib[q1, a, i], psi[q2, a, i], psib[q3, b, j], 
  psi[q3 + q1 - q2, b, j]]];
\end{verbatim}
In this expression Dirac $\gamma$ matrices are represented by the function \tw{diracM} where the first argument denotes 
the momentum contracted with the $\gamma$-matrix and the other two represent the indices of the matrix. 
The wave-function renormalization is denoted by \tw{Z} and the regulator function by \tw{Rs}.

\begin{figure}
 \begin{center}
  \includegraphics[width=0.5\textwidth]{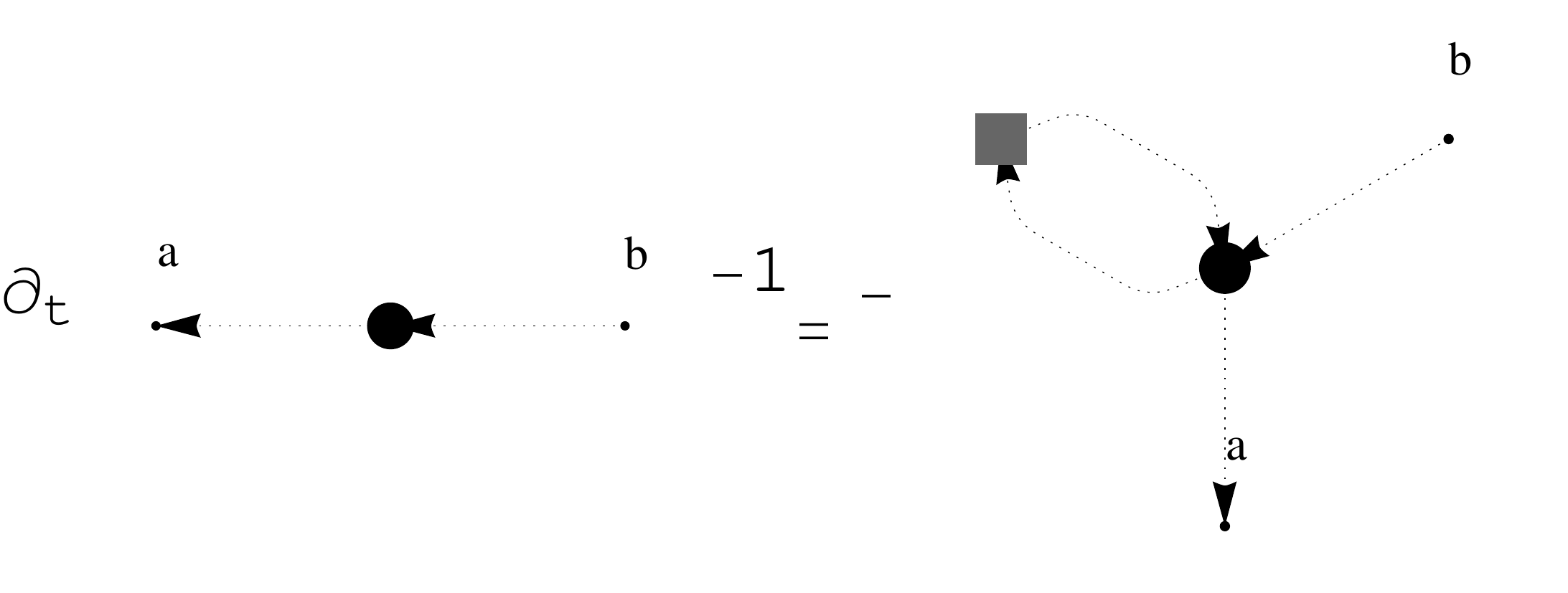}\\
  \includegraphics[width=0.95\textwidth]{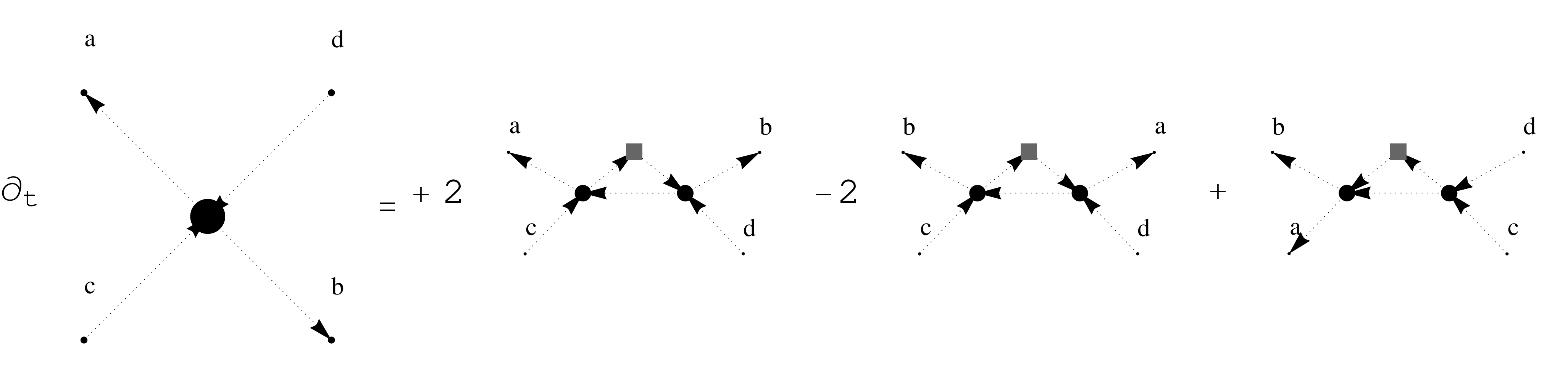}
  \caption{\label{fig:GNFlows}Flows of the two- and the four-point functions of the Gross-Neveu model.}
 \end{center}
\end{figure}

The flows of the two- and four-point functions can be obtained from\footnote{Note that for illustration purposes we directly use the 
physical instead of the symbolic action.}
\begin{verbatim}
twoPoint = doRGE[actionGN, {psib, psi}];
fourPoint = doRGE[actionGN, 
 {{psib, i1}, {psib, i2}, {psi, i3}, {psi, i4}}];
\end{verbatim}
The results are depicted in \fref{fig:GNFlows}.

We now define the Feynman rules. From the action we obtain the algebraic expressions for 
the two- and the four-point functions
\begin{verbatim}
FR2Point = getFR[actionGN, {psib[p1, a, i], psi[p2, b, j]}]

--> delta[flavor, i, j] deltam[-p1 + p2] diracM[p1, a, b]
  Rs[p1^2, k] + delta[flavor, i, j] deltam[-p1 + p2]
  diracM[p1, a, b] Z[p1^2]
\end{verbatim}
and
\begin{verbatim}
FR4Point = getFR[actionGN, {psib[p1, a, i], psib[p2, b, j],
  psi[p3, c, m], psi[p4, d, l]}]

--> -((gb delta[dirac, a, d] delta[dirac, b, c] 
  delta[flavor, i, m] delta[flavor, j, l]
  deltam[-p1 - p2 + p3 + p4])/Nf) + 
 (gb delta[dirac, a, c] delta[dirac, b, d] 
  delta[flavor, i, l] delta[flavor, j, m] 
  deltam[-p1 - p2 + p3 + p4])/Nf
\end{verbatim}
We use the expression for the four-point function directly and infer the form of the propagator from that of the two-point function:
\begin{verbatim}
P[psi[p1_, a_, i_], psib[p2_, b_, j_], explicit -> True] := 
 delta[flavor, i, j] diracM[p1, a, b]/
  p1^2/(Z[p1^2] + Rs[p1^2, k]);
V[psib[p1_, a_, i_], psib[p2_, b_, j_], psi[p3_, c_, m_], 
   psi[p4_, d_, l_], explicit -> True] = 
  FR4Point/deltam[-p1 - p2 + p3 + p4] // Expand;
\end{verbatim}
Recall that \textit{DoFun} implicitly assumes a momentum conserving $\delta$ distribution to be present in its output.
Furthermore we need to define the regulator insertion which we choose to be diagonal in flavor space:
\begin{verbatim}
dR[psib[p1_, a_, i_], psi[p2_, b_, j_], explicit -> True] := 
  diracM[p1, a, b] delta[flavor, i, j] dRs[p1^2, k];
\end{verbatim}

Before we derive the algebraic form of the equations let us introduce dummy index names for Dirac (\tw{a}, \tw{b}, and so on) 
and flavor indices (\tw{i}, \tw{j}, and so on):
\begin{verbatim}
addIndices[{{flavor, {i, j, k, l, m, n}},
 {dirac, {a, b, c, d, e, f,  g, h}}}];
\end{verbatim}
We also define some rules for calculations and convenient replacements later:
\begin{verbatim}
dimRules = {dim[flavor] :> Nf, dim[dirac] :> dgamma};
diracRules = {diracM[-p_, a_, b_] :> -diracM[p, a, b], 
   diracM[p_, a_, a_] :> 0, 
   diracM[q_, a_, b_] diracM[q_, b_, c_] :> q^2 delta[dirac, a, c], 
   diracM[q_, b_, a_] diracM[q_, c_, b_] :> q^2 delta[dirac, a, c], 
   diracM[q_, a_, b_]^2 :> q^2 delta[dirac, a, a]};
\end{verbatim}
The latter set of rules is useful for the evaluation of traces of products of Dirac matrices.

Now we have all the ingredients at hand which are required to derive the algebraic expression for the
two-point function with \tw{getAE}:
\begin{verbatim}
twoPointAlg = Plus @@ getAE[twoPoint,
 {{psi, i2, p1, b, j}, {psib, i1, p2, a, i}}] // Expand;
\end{verbatim}
This is the flow equation of $\frac{\de^2 \Gamma_k}{\de \bar{\psi}^b_i(p_1)\de \psi^b_j(p_2)}$.

For simplicity, we restrict ourselves to the point-like limit, i.~e. we set the external momenta to zero.
To simplify the expression \tw{twoPointAlg}, we repeatedly apply the function \tw{integrateDeltas} as well as the 
rules \tw{diracRules} until the output does not change any further:
\begin{verbatim}
FixedPoint[integrateDeltas[#] /. diracRules &, 
 twoPointAlg /. {p1 -> 0, p2 -> 0}]

--> (gb delta[flavor, i, j] diracM[q1, a, b] dRs[q1^2, k])/
 (Nf q1^2 (Rs[q1^2, k] + Z[q1^2])^2)
\end{verbatim}
Translated into the standard textbook notation, this expression reads
\begin{align}
\frac{\overset{\rightarrow}{\de} }{\de \bar{\psi}^a_i(p_1)}\partial_t \Gamma_k \frac{\overset{\leftarrow}{\de}}{\de \psi^b_j(p_2)}\Bigg|_{\psi=\bar{\psi}=0}=\de_{ij}\int \ddotp{q} \frac{\bar{g}}{N_f} q_\mu \gamma_\mu^{ab} \frac{\partial_t R_k(q^2)}{q^2(Z_k(q^2)+R_k(q^2))^2}.
\end{align}
As expected, this expression vanishes in the point-like limit since 
the integrand is anti-symmetric in the four-momentum $q_{\mu}$.

The flow equation of  the four-point function can be derived along the lines of the derivation of the two-point function.
For convenience, however, we set the external momenta to zero right from the beginning:
\begin{verbatim}
fourPointAlg = Plus @@ getAE[fourPoint,
  {{psib, i1, 0, a, i}, {psib, i2, 0, b, j},
   {psi, i3, 0,  c, l}, {psi, i4, 0, d, m}}];
\end{verbatim}
Applying the same simplification procedure as above yields
\begin{verbatim}
fourPointAlgSim=Simplify[FixedPoint[integrateDeltas[#] /.
 diracRules &, fourPointAlg]]

--> (1/(Nf^2 q1^2 (Rs[q1^2, k] + Z[q1^2])^3))2 gb^2 
  (-2 + dgamma Nf) (delta[dirac, a, d] delta[dirac, b, c]
  delta[flavor, i, m] delta[flavor, j, l] - 
 delta[dirac, a, c] delta[dirac, b, d] delta[flavor, i, l]
  delta[flavor, j, m]) dRs[q1^2, k]
\end{verbatim}
This is the right-hand side of the flow equation of the four-point function 
$\frac{\de^4 \Gamma_k}{\de \bar{\psi}_i^a\de \bar{\psi}^b_j\de \psi _l^c \psi^d_m}$. 
There are only two terms left. Thus, contributions from the three different diagrams must have canceled each other.
Since we would like to compute the flow of the coupling $\bar{g}$, we have to project the output \tw{fourPointAlgSim} onto our ansatz~\eref{eq:ansatzGN}.
To this end we contract the output  \tw{fourPointAlgSim} with the tensor-structure of the four-fermion interaction given by 
\begin{verbatim}
V[psib[p1, a, i], psib[p2, b, j],
   psi[p3, c, l], psi[p4, d, m], explicit -> True]/gb
\end{verbatim}
This yields the flow equation for $\bar{g}$:
\begin{verbatim}
fourPointAlgSimProj = FixedPoint[integrateDeltas[#] /.
  diracRules &, V[psib[p1, a, i], psib[p2, b, j],
   psi[p3, c, l], psi[p4, d, m], explicit -> True]/gb
  fourPointAlgSim] /. dimRules;
lhs = integrateDeltas[
  V[psib[p1, a, i], psib[p2, b, j], psi[p3, c, l],
   psi[p4, d, m], explicit -> True] 
  V[psib[p1, a, i], psib[p2, b, j], psi[p3, c, l],
   psi[p4, d, m], explicit -> True]/gb^2] /. dimRules;
\end{verbatim}
\tw{lhs} corresponds to the projection of \tw{fourPointAlgSim} on the left-hand side. Thus, the flow equation for the four-fermion interaction
$\bar{g}$ reduces to the following simple expression
\begin{verbatim}
beta = fourPointAlgSimProj/lhs // Simplify

--> -((2 gb^2 (-2 + dgamma Nf) dRs[q1^2, k])/
 (Nf q1^2 (Rs[q1^2, k] + Z[q1^2])^3))
\end{verbatim}

Let us rewrite this expression with the aid of the following replacement rules: 
\begin{verbatim}
dimLessRulesGN = {Rs[q1^2, k] :> r[y] Z[q1^2], gb -> g k^(2 - d), 
   dRs[q1^2, k] :> -2 y Z[q1^2] r'[y], q1 :> Sqrt[y] k, 
   dq1 :> k^2/2/q1};
\end{verbatim}
Here, \tw{r[y]} denotes the dimensionless regulator function, \tw{g} the dimensionless coupling constant and \verb|y=q1^2/k^2|. \tw{dq1} is the integral 
measure. Theses rules allow us to obtain a flow equation in terms of dimensionless quantities.
The integral over the angular part is
\begin{verbatim}
angleInt[d_] := 4v[d];
\end{verbatim}
where \tw{v[d]} represents $v_d=(\pi^{d/2} 2^{d + 1} \Gamma(d/2))^{-1}$.
Next, we rewrite the flow equation in terms of dimensionless quantities:
\begin{verbatim}
betaDimLess = 
  angleInt[d] dq1 q1^(d - 1) beta //. dimLessRulesGN// Expand;
Simplify[betaDimLess, Assumptions -> Element[d, Integers]]

--> (8 g^2 k^(2 - d) (-2 + dgamma Nf) y^(-1 + d/2) v[d] r'(y))/
 (Nf (1 + r[y])^3 Z[k^2 y]^2)
\end{verbatim}
Translated into standard textbook notation this expression reads
\begin{align}
  (d_\gamma N_f-2)k^{2-d}\frac{8 g^2 v_d}{N_f}\int_0^\infty dy\frac{y^{\frac{d}{2}-1} r'(y)}{ (1+r(y))^3 Z(k^2 y)^2}.
\end{align}

To obtain the final expression for the flow equation we also have to manipulate the left-hand side. It reads
\begin{align}
-&\frac{\overset{\rightarrow}{\de}^2 }{\de \bar{\psi}_i^a(p)\de \bar{\psi}^b_j(q)\de}\partial_t \Gamma_k \frac{\overset{\leftarrow}{\de}^2 }{\psi _l^c(r) \psi^d_m(s)}\Bigg|_{\psi=\bar{\psi}=0}=\nnnl
 =&\left((2-d-2\eta_\psi)g+\partial_t g\right)\,k^{2-d} \times\nnnl
&\times\frac1{N_f}\left(\de_{il}\de_{jm}\de^{ac}\de^{bd} - \de_{im}\de_{jl}\de^{ad}\de^{bc}\right)(2\pi)^d\delta(p+q+r+s)\label{eq:lhs4psi}
\end{align}
where $\eta_\psi=-\partial_t \ln Z$ is the anomalous dimension associated with the field $\psi$.
Note the minus sign in front of the derivatives which appears due to our conventions, see \eref{eq:effActions}. 
In the point-like limit the projection yields
\begin{align}
\left((2-d-2\eta_\psi)g+\partial_t g\right)\,k^{2-d} (2\pi)^d\delta(0).
\end{align}
Combining the left- and right-hand sides we obtain the flow equation for $g$:
\begin{align}
\partial_t g=(d-2+2\eta_\psi)g+\left(d_\gamma -\frac{2}{N_f}\right)8v_d\int_0^\infty dy\,\frac{ y^{\frac{d}{2}-1} r'(y)}{ (1+r(y))^3 Z^2}g^2.
\end{align}
In the large $N_f$-limit this reduces to the result found in Ref.~\cite{Braun:2010tt}.

Of course, the six-point function can be also derived with \DoFun. From the symbolic expression
\begin{verbatim}
sixPoint = doRGE[actionGNSymbolic, 
 {psib, psib, psib, psi, psi, psi}];
\end{verbatim}
(which yields 45 terms as can be checked with \tw{countTerms@sixPoint})
we obtain the algebraic expression:
\begin{verbatim}
sixPointAlg =  Plus @@ getAE[sixPoint,
{{psib, i1, 0, a1, j1}, {psib, i2, 0, a2, j2}, {psib, i3, 0, a3, j3},
 {psi, i4, 0, a4, j4}, {psi, i5, 0, a5, j5}, {psi, i6, 0, a6, j6}}];
\end{verbatim}
Simplifying the result reveals that each term contains an odd number of $\slashed{q}$ and thus the flow of the six-point function
vanishes upon integration in the point-like limit as expected.

To illustrate the flexibility of the developed formalism we would like to
mention how easy it is to extend our study to non-vanishing external momenta.
With \DoFun, we only need to change a few arguments and define some additional rules for the handling of the Dirac algebra. 
We demonstrate this for the flow equation of the four-point function. In short, we perform the following steps: Derive the algebraic RG equation,

\begin{verbatim}
fourPointAlgFinMom = 
  Plus @@ getAE[
    fourPoint, {{psib, i1, p1, a, i}, {psib, i2, p2, b, j},
     {psi, i3, p3, c, l}, {psi, i4, p4, d, m}}];
\end{verbatim}
project it using
\begin{verbatim}
fourPointAlgProj = 
 FixedPoint[integrateDeltas[#] /. diracRules &, 
   V[psib[p1, a, i], psib[p2, b, j], psi[p3, c, l], psi[p4, d, m], 
      explicit -> True]/gb fourPointAlgFinMom] /. dimRules;
betaFinMom = fourPointAlgFinMom/lhs;
\end{verbatim}
and then make it dimensionless,
\begin{verbatim}
betaFinMomDimLess = 
 angleInt[d] dq1 q1^(d - 1) betaFinMom //. dimLessRulesGN // Expand;
\end{verbatim}
For simplicity, we finally present the result for the large-$N_{\rm f}$ limit:
\begin{verbatim}
betaDimLessLargeNf = Limit[betaDimLess, Nf -> \[Infinity]];
Simplify[betaDimLessLargeNf, Assumptions -> Element[d, Integers]]

--> -(4 g^2 k^(2 - d) y^(-1 + d/2)
     diracM[k Sqrt[y], c$180535, f$180535] v[
     d] ((p1 + p3 + k Sqrt[y])^2 diracM[-p2 - p3 - k Sqrt[y], 
        f$180535, 
        c$180535] (Rs[(p1 + p3 + k Sqrt[y])^2, k] + 
         Z[(p1 + p3 + k Sqrt[y])^2]) + (p2 + p3 + 
         k Sqrt[y])^2 diracM[-p1 - p3 - k Sqrt[y], f$180535, 
        c$180535] (Rs[(p2 + p3 + k Sqrt[y])^2, k] + 
         Z[(p2 + p3 + k Sqrt[y])^2])) Derivative[1][r][
     y])/((p1 + p3 + k Sqrt[y])^2 (p2 + p3 + k Sqrt[y])^2 (1 + 
      r[y])^2 (Rs[(p1 + p3 + k Sqrt[y])^2, k] + 
      Z[(p1 + p3 + k Sqrt[y])^2]) (Rs[(p2 + p3 + k Sqrt[y])^2, k] + 
      Z[(p2 + p3 + k Sqrt[y])^2]) Z[k^2 y])
\end{verbatim}
This result can be further simplified and used in numerical calculations.

\section{Summary}
In this paper we have presented \DoFun, a \textit{Mathematica} application which allows to derive both Dyson-Schwinger equations
and functional RG equations starting from a given action. \DoFun\ is based on \textit{DoDSE}, see Ref.~\cite{Alkofer:2008nt}, which was 
limited to the derivation of DSEs in symbolic form. \DoFun\ goes beyond the symbolic form and provides explicit expressions for the integrals.
Apart from the functions used for the actual derivation of the equations, we have included several additional helpful tools, e.~g. for dealing with the Kronecker delta and for the derivation of Feynman rules.

We have demonstrated the usage of \DoFun\ by means of a scalar $O(N)$ field theory and the Gross-Neveu model in a purely fermionic description.
In particular, $O(N)$ symmetric scalar field theories have been studied in great detail in the literature, see e.~g. Refs.~\cite{Tetradis:1993ts,Litim:2002cf,Braun:2008sg,Benitez:2009xg,Litim:2010tt}.
Although these theories are well suited to demonstrate the usage of \DoFun, our main goal was to provide a tool which facilitates the derivation
of functional equations for even more involved theories, such as QED or QCD, where the tensor structure leads to additional complications. 
Furthermore, the numbers of terms grow considerably in gauge theories when 
higher $n$-point functions are taken into account. Therefore we hope that \DoFun\ proves to be a valuable tool in particular for functional studies of gauge theories 
and helps to push them to new limits.

We would like to encourage users of \DoFun\ to actively communicate their experiences with it in order to help us to further improve this application. 
In particular, bug reports are most welcome.

\section*{Acknowledgments}
JB and MQH are grateful for useful discussions with L. Fister, H. Gies, L. M. Haas, J. M. Pawlowski and A. Wipf. The authors would like to thank J. M. Pawlowski and A. Wipf for critical comments on the manuscript.
MQH acknowledges support by DFG Gi 328/1-4. JB and MQH
acknowledge support by the DFG research training group GRK 1523/1.

\appendix

\section{How to include Grassmann fields}
\label{app:fermions}

Implementing the anti-commuting nature of fermionic fields into a \textit{Mathematica} program like \textit{DoFun} 
requires the use of a non-commuting multiplication operator. 
This was one reason for introducing the \texttt{op}-function as described in section \ref{ssec:language}. 
It allows to properly take into account the effects of changing the order of the fields. Internally all anti-commuting fields are called either \tw{fermion} or \tw{antiFermion}. This also includes, e.~g., ghost fields, which are not fermions, but anti-commute as well.
The corresponding test functions are \tw{fermionQ} and \tw{antiFermionQ}, while \tw{grassmannQ} yields true in either case. 

\textit{DoFun} recognizes fermions via the definition of the two-point functions:
If any two-point function in the action has two fields with different names, \textit{DoFun} considers the first one as a fermion field
and the second one as its anti-field. In case of complex fields or a mixing of bosonic fields at the level of two-point functions
this can be suppressed with the option \texttt{specificFieldDefinitions}, see section \ref{ssec:actions}. \DoFun\ also 
allows to define all fields as bosons or fermions without specifying two-point functions.
An example for a fermionic action is given by the Gross-Neveu model\footnote{In this section $\psi$ and $\bar{\psi}$ 
correspond to Grassmann fields, while $\Phi$ is a bosonic field.}:
\begin{verbatim}
actionGNSymbolic={{psi, psib}, {psib, psib, psi, psi}};
\end{verbatim}
Here, the canonical order was used where anti-fermions (\tw{psib} for $\bar{\psi}$) appear always to the
left of fermions (\tw{psi}) in $n$-point functions.
This is due to the fact that we employ the following definitions for Grassmann derivatives:
\begin{align}
& \frac{\delta}{\delta \psi} M:= M\frac{\overset{\leftarrow}{\delta}}{\delta \psi}, & \frac{\delta}{\delta \bar{\psi}}M := \frac{\overset{\rightarrow}{\delta}}{\delta \bar{\psi}}M.
\end{align}
Consequently, derivatives with respect to fermions are assumed to act from the right whereas
derivatives with respect to anti-fermions are assumed to act from the left. For fermions, the differentiation rules of \eref{eq:derivatives} have to be 
adapted as follows:
\begin{subequations}\label{eq:derivativesGrassmann}
\begin{align}
\frac{\delta}{\delta\bar{\psi}_{i}}D^{jk}_{\Phi\Phi,J} & =\frac{\delta}{\delta\bar{\psi}_{i}}\left(\frac{\delta^{2}\Gamma}{\delta\Phi_{j}\delta\Phi_{k}}\right)^{-1}=D^{jm}_{\Phi\Phi,J}\Gamma^{imn}_{\bar{\psi}\Phi\Phi,J}D^{nk}_{\Phi\Phi,J}, \\
\frac{\delta}{\delta\psi_{i}}D^{jk}_{\Phi\Phi,J} & =\left(\frac{\delta^{2}\Gamma}{\delta\Phi_{j}\delta\Phi_{k}}\right)^{-1}\frac{\overset{\leftarrow}{\delta}}{\delta \psi}=D^{jm}_{\Phi\Phi,J}\Gamma^{mni}_{\Phi\Phi\psi,J}D^{nk}_{\Phi\Phi,J}, \\
\frac{\delta}{\delta\bar{\psi}_{i}}\Gamma^{j_{1}\ldots j_{n}}_{\Phi\ldots\Phi,J} & =-\frac{\delta\Gamma}{\delta\bar{\psi}_{i}\delta\Phi_{j_{1}}\ldots\delta\Phi_{j_{n}}}=\Gamma^{ij_{1}\ldots j_{n}}_{\delta\bar{\psi}\Phi\ldots\Phi,J},\\
\frac{\delta}{\delta\psi_{i}}\Gamma^{j_{1}\ldots j_{n}}_{\Phi\ldots\Phi,J} & =-\frac{\delta\Gamma}{\delta\Phi_{j_{1}}\ldots\delta\Phi_{j_{n}}\delta\psi_{i}}=\Gamma^{ij_{1}\ldots j_{n}}_{\Phi\ldots\Phi\psi,J}.
\end{align}
\end{subequations}

To alleviate the readability of expressions, propagators are defined with fermions and anti-fermions exchanged, i.~e. a propagator $D^{\bar{\psi}\psi}_{ij}(p^2)$ is in \DoFun\ correctly represented by \texttt{P[\{psi,i\}, \{psib,j\}]}. As a consequence, the connecting fields of a vertex and a propagator are the same even for fermions, whereas normally one would have a fermion and its anti-fermion.
For example, in the two-point DSE of the Gross-Neveu model we encounter the 
following tadpole-like diagram where the leg \verb|{psib,r1}| of the vertex is attached to the leg \verb|{psib,r1}| of the propagator:
\begin{verbatim}
op[S[{psib, i2}, {psib, r1}, {psi, s1}, {psi, i1}],
 P[{psi, s1}, {psib, r1}]]
\end{verbatim}
We choose this notation to make it easier to connect the propagators and vertices visually when seeing this output.

Using left- and right-derivatives also has consequences for the definition of vertices. For instance, a four-fermion interaction has the form
\begin{align}
 \Gamma^{ijkl}_{\bar{\psi}\bar{\psi}\psi\psi}=-\frac{\de^4 \Gamma}{\de \bar{\psi}_i \de \bar{\psi}_j \de \psi_k \de \psi_l}\Big|_{\bar{\psi}_{phys},\psi_{phys}}:=-\frac{\overset{\rightarrow}{\delta}^2 }{\delta \bar{\psi}_i \de \bar{\psi}_j }\Gamma\frac{\overset{\leftarrow}{\delta}^2}{\de \psi_k \de \psi_l} \Big|_{\bar{\psi}_{phys},\psi_{phys}}.
\end{align}
Thus, the indices in $\Gamma^{ijkl}_{\bar{\psi}\bar{\psi}\psi\psi}$ do not reflect the order in which the derivatives are performed, 
but rather the order in which the derivatives appear. In this example this means that the differentiation with respect to $\bar{\psi}_j$ ($\psi_k$) has to be performed 
before that with respect to $\bar{\psi}_i$ ($\psi_l$). This convention is employed for bare and dressed 
vertices, \tw{S} and \tw{V}, respectively, and also in \tw{getFR}. To derive the expression for the bare four-point 
function $S^{ijkl}_{\bar{\psi}\bar{\psi}\psi\psi}$ from a given action one uses
\begin{verbatim}
getFR[action, {psib[p1,i], psib[p2,j], psi[p3,k], psi[p4,l]}]
\end{verbatim}
The required reordering of anti-fermions is done automatically by \tw{getFR}.

An exception to that rule for fermion ordering are the derivative arguments in \tw{doDSE}. Here, 
the order corresponds exactly to the order in which the derivatives are performed. This is due to the fact 
that the form of a DSE depends on which field is attached to the bare vertex. For the ghost-gluon vertex of Landau gauge Yang-Mills theory this is explicitly demonstrated in ref.~\cite{Alkofer:2008nt}.
For example, the DSE of the 
four-point function $\Gamma^{ijkl}_{\bar{\psi}\bar{\psi}\psi\psi}$ can be obtained from
\begin{verbatim}
doDSE[action, {{psib, j}, {psib, i}, {psi, k}, {psi, l}}]
\end{verbatim}
but also from
\begin{verbatim}
doDSE[action, {{psi, k}, {psi, l}, {psib, j}, {psib, i}}]
\end{verbatim}
Other variations also exist. The important point is that the first derivative has to be taken either with respect to $\psi_k$ or $\bar{\psi}_j$. For RGEs the same rule applies.

Now we turn to the minus signs arising from the anti-commutativity of fermions. As we use superfields
which may have bosonic as well as fermionic entries, some care is required in the derivation.  
For DSEs the minus signs for fermionic loops arise immediately from the ordering of the fields at the end of the calculation. 
An example in ref. \cite{Alkofer:2008nt} nicely demonstrates this. However, the currently employed algorithm is not infallible: There are known examples with a wrong sign, but the lowest diagram affected is a two-loop diagram of a three-point function. In this case the sign needs to be corrected by hand.

For RGEs we employ a completely different algorithm and up to now we are not aware of any shortcomings.
In contrast to DSEs the order for the field derivatives does not affect the result. Therefore we are free to 
put them into canonical order: anti-fermions always appear to the left of fermions to the left of bosons.
The order between fields of the same 'type' is not changed. Hence \tw{doRGE} \textit{always} uses this canonical order to derive RGEs.
This ordering is in agreement with our choice given in \eref{eq:flowEqLog}. 
For the differentiation itself we ignore any minus signs that would arise by dragging one Grassmann derivative to the left or to the right of 
another one.
The minus signs from passing two Grassmann fields of the external derivatives are taken into account at the end. 
As far as the internal fields are concerned, the nature of the fields (bosonic or Grassmann-valued) is of no importance
since after setting the sources to zero all propagators, vertices and regulator insertions have zero Grassmann number and can be freely commuted. 
In fact, we find that all minus signs that would arise if we took into account the Grassmann numbers of internal fields cancel each other. 
Thus, the only signs we have to worry about stem from interchanging the external field derivatives. Those are corrected
in a last step by comparing their order with that of the original canonically ordered derivatives.

\section{Version overview of \textit{DoFun}}
\label{app:versions}

The predecessor of \DoFun\ is \textit{DoDSE}, see Ref.~\cite{Alkofer:2008nt}. Subsequently smaller updates have been 
made available. With version $2.0$ the package \textit{DoDSE} became part of the application \DoFun\ 
together with the new packages \textit{DoAE} and \textit{DoFR}. The current version is $2.0$.

The following list gives a short overview over all publicly available versions:
\begin{itemize}
 \item \textit{DoDSE 1.0} (Aug. 15, 2008): first publicly available version
 \item updates \textit{DoDSE 1.1}, \textit{1.2}, \textit{1.2.1}, \textit{1.2.2}, \textit{1.2.3} (Feb. 9, 2010)
 \item \textit{DoFun 2.0} (Feb. 25, 2011): contains \textit{DoDSERGE} (formerly \textit{DoDSE}), \textit{DoAE} and \textit{DoFR}
\end{itemize}

Further updates will be made available at
\begin{center}\url{http://theorie.ikp.physik.tu-darmstadt.de/~mqh/DoFun}.\end{center}
If an Internet connection is available, \DoFun\ will automatically notify the user about available updates.



\setstretch{0.5}

\bibliographystyle{utphys_mod}
\bibliography{literature_RGE-Derivation}

\end{document}